\documentclass[graybox, nosecnum]{svmult}

\usepackage{mathptmx}       

\usepackage{helvet}         
\usepackage{courier}        
\usepackage{type1cm}        
\usepackage{comment}
\usepackage{nameref}
\usepackage{makeidx}         
\usepackage{graphicx}        
\usepackage{multicol}        
\usepackage[bottom]{footmisc}
\usepackage{hyperref}        
\usepackage{soul}            
\usepackage{siunitx}[=2021-04-09]
\usepackage{array,multirow}
\hypersetup{colorlinks=true,urlcolor=blue}
\usepackage[square,numbers]{natbib}

\newcommand{\athena}{{\it Athena}}

\newcommand{\claire}{{\it CLAIRE}}

\makeindex             
                       
\begin{document}

\title*{Laue and Fresnel lenses}
\author{Enrico Virgilli\thanks{corresponding author}, Hubert Halloin, 
and Gerry Skinner}
%

\institute{
Enrico Virgilli \at Istituto Nazionale di Astrofisica INAF-OAS, Via Piero Gobetti, 93/3, 40129 Bologna - Italy\\ \email{enrico.virgilli@inaf.it}
\and Hubert Halloin\at  Université de Paris, CNRS, Astroparticule et Cosmologie, F-75013 Paris, France\\ \email{hubert.halloin@apc.in2p3.fr}
\and Gerald Skinner\at University of Birmingham, Birmingham B15 2TT, UK\\ \email{gerald.k.skinner@gmail.com}
}
\maketitle
\section*{Abstract}
\label{abstract}
The low-energy gamma-ray domain is an important window 
for the study of the high energy
Universe. Here matter can be observed in extreme 
physical conditions and during powerful
explosive events. However, observing gamma-rays 
from faint sources is extremely challenging with 
current instrumentation. With techniques used 
at present collecting more signal requires 
larger detectors, leading to an increase in 
instrumental background. For the leap in sensitivity 
that is required for future gamma-ray missions use
must be made of flux concentrating telescopes. 
Fortunately, gamma-ray optics such as Laue or
Fresnel lenses, based on diffraction,  make 
this possible. Laue lenses 
work with moderate focal
lengths (tens to a few hundreds of metres), but provide only 
rudimentary imaging capabilities. 
On the other hand,
Fresnel lenses offer extremely good  imaging, 
but with a very 
small field of view and a requirement for 
focal lengths $\sim$10$^8$ m. 
This chapter presents the basic concepts
of these optics and describes their working 
principles, their main properties and  
some feasibility studies already conducted.

\section{Introduction}
\label{introduction}

The `low-energy' gamma-ray band from $\sim100$~keV to a few tens of MeV is of crucial importance in the understanding of many astrophysical processes. It is the band in which many astrophysical systems emit most of their energy.
It also contains the majority of gamma-ray lines from the decay 
of radioactive nuclei associated 
with synthesis of the chemical elements
and also the 511~keV line tracing the annihilation of positrons.
However, observations at these energies are constrained in ways that those at lower and higher energies are not. 
At  lower energies grazing incidence optics enable true focusing of the incoming radiation, forming images and concentrating power from compact sources onto a small detector area.  At higher energies the pair production process allows the direction of the incoming photon to be deduced. 
But in the low energy gamma-ray band grazing incidence optics are impractical (the graze angles are extremely small) and the dominant Compton interaction process provides only limited directional information. 

Detector background due to particle interactions and to photons 
from outside the region of interest is a major problem in 
gamma-ray astronomy. A  large collecting area is essential 
because the fluxes are low, but unless a means is found to 
concentrate the radiation this implies a large detector and 
hence a lot of background. Shielding helps but it is 
imperfect and the materials in the shield themselves 
produce additional background. 
If the flux from a large collecting area $A$ can 
be concentrated with efficiency $\eta$ onto a small 
area $A_d$ of detector then for background-dominated 
observations there is an advantage in sensitivity of  
$\sqrt{\eta A/A_d}$ compared with a `direct-view'  
instrument of area $A$ having the same background 
per unit area, energy band and observation time. 

 At energies where grazing incidence optics are not viable, 
 only two technologies are available for concentrating 
 gamma-rays. Both use diffraction. Laue lenses use diffraction 
 from arrays of crystals while Fresnel lenses utilise diffraction 
 from manufactured structures. 
 Both type of lens can provide a high degree of concentration 
 of flux from a compact on-axis source.
 Fresnel lenses provides true 
imaging, albeit with chromatic aberrations, whereas the 
Laue lens is a `single-reflection' optic, where the 
off-axis aberrations are severe.

MeV astrophysics is now eagerly waiting for the 
launch of NASA's COSI mission \cite{Tomsick2019}
which is scheduled for 2025.
This will be a survey mission with a large field 
of view - and consequently a relatively high background.
Nevertheless, {\textcolor{green}{COSI}} is expected to be a factor of 10 more 
sensitive than {\textcolor{green}{Comptel}} the pioneering instrument for 
the MeV gamma-ray range \cite{shoenfelder93}. A focusing 
telescope using the techniques discussed here may improve the sensitivity for the study of individual sources by 
another large factor,
allowing studies not possible with scanning, high-background, instruments.

Laue lenses and Fresnel 
lenses are discussed separately below.


\section{Laue lenses}
\begin{figure}[!t]
  \center
     \includegraphics[scale=1,keepaspectratio=true]{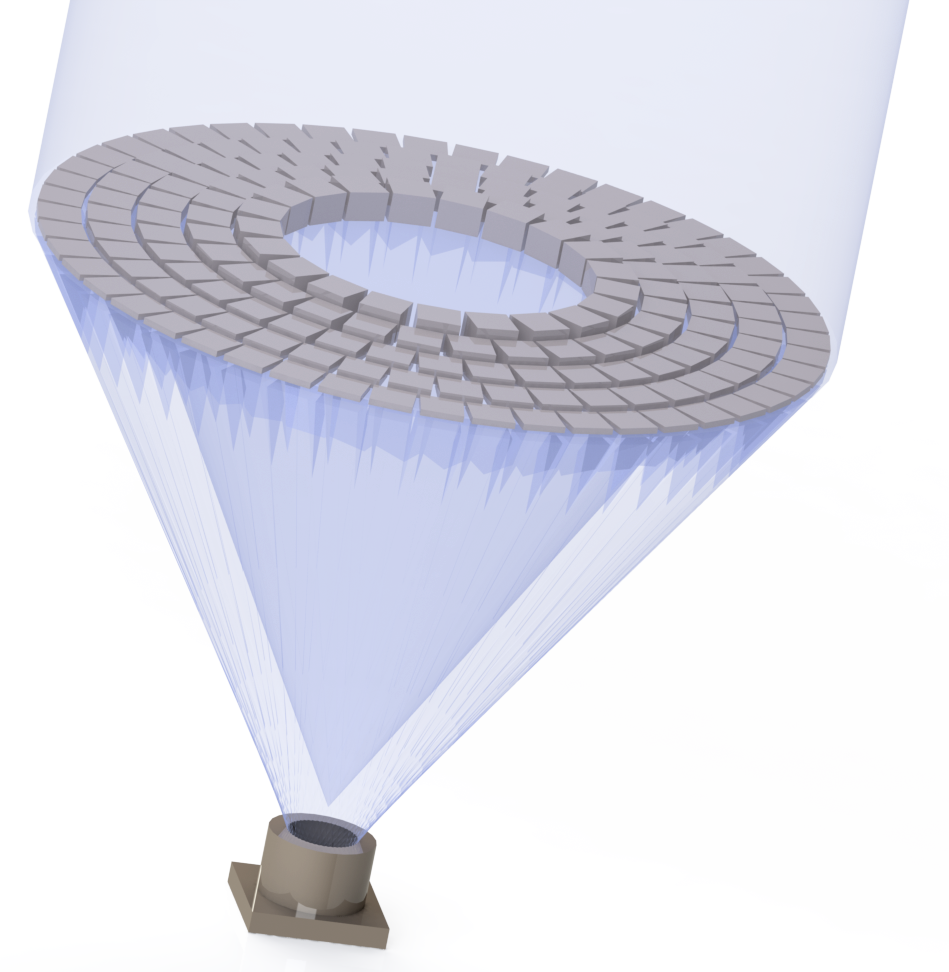}
  \caption{A gamma-ray lens, based on diffraction 
  from crystals in the Laue (transmission) geometry. Crystal tiles are  oriented in order to diffract the radiation towards the common focus. In this drawing 5 concentric rings of crystals are shown as an example. Crystals can be arranged over a spherical cap but the lens can also be planar and other radial distributions are possible. Crystals at the same radius from the axis 
  will have the same orientation with respect to the incident  radiation but the orientation will change with radius.}
  \label{fig:LLconcept1}
\end{figure}

The concept of a Laue telescope is shown in Figure \ref{fig:LLconcept1}. The essential element is a `{\textcolor{green}{Laue Lens}}' containing a large number of high-quality crystals.
Each crystal must be correctly oriented to diffract 
radiation 
in a narrow spectral range from a distant source 
towards a detector located at a common 
focus behind the lens.
The crystals are used in the Laue-mode (transmission) 
since the very small diffraction angles make it impractical 
to rely on surface reflections. Excellent reviews of previous 
work on this topic can be found in \cite{frontera10a} and 
\cite{Smither2014}.

The term `Laue lens´ is actually a misnomer and it would be 
more correct to refer to a Laue-mirror. Such a `lens' relies 
on the mirror \emph{reflection} of gamma-rays from the lattice
planes in the crystals. The reflective power of the electrons 
bound in atoms in a single lattice plane is very small, but 
the power increases with the square of the number of 
planes - or electrons - acting coherently. 
\subsection{Laue lenses basic principles: Bragg's law}
The requirement for coherent diffraction is both the strength and
the weakness of Laue lenses. It provides for the possibility of high reflectivity, but at the same time it
imposes  a strict dependence of the diffraction angle, $\theta_B$, on
the wavelength, $\lambda$, (or the energy, $E$) of the radiation. This dependence is
expressed by {\textcolor{green}{Bragg's law}}:
\begin{equation}
  \sin\theta_B = \ n \frac{\lambda}{2 d_{hkl}} \hspace{3em}{\rm with: }\hspace{1em} n=1,2,3,... \ 
\label{eq:t_bragg}
\end{equation}
where $n$ is the diffraction order, $d_{hkl}$ is the spacing of the crystal lattice planes and
$\lambda$ is the wavelength of the gamma-rays. The first order ($n$ = 1) contributions
are by far the most significant. For energies which concern us here
the Bragg angles are always small $(\simeq 1^{\rm o} )$, so we can set 
$\tan\theta = \sin\theta = \theta$.  
In the following we shall often prefer to speak in terms of energy, $E$, rather than
wavelength. Bragg's law then takes the form:
\begin{equation}
  \sin\theta_B = \ n \frac{h c}{2 d_{hkl} E} \ ,
\label{eq:E_bragg}
\end{equation}
where $h$ is Planck's constant and $c$ the velocity of light.   
($\lambda$ (Å) $= 12.39 / E$ (keV)).

From the Bragg equation for first diffraction order, with simple geometrical considerations it can be shown that there is a relation between the distance $r_i$ at which the crystal is positioned  with respect to the Laue lens axis and the diffracted energy E$_i$:
\begin{equation}
E_i = \frac{hc F}{d_{hkl} r_i},   
\label{e_r}
\end{equation}
where F is the focal length  of the Laue lens. 

Eq.~\ref{e_r} shows that for a given focal length, if crystals with a fixed $d_{hkl}$ are used, those placed closer to the lens axis are dedicated to the highest energies while those positioned further away from the axis diffract the lower energies. Consequently, at a given focal length there is a direct link between the energy band of a Laue 
lens and its spatial extent, r$_{in}$ to r$_{out}$.
For a narrow band Laue lens, with the whole surface optimised for a single energy, it would be necessary to arrange that $d_{hkl}$ varies in the radial 
direction such that ${d_{hkl} r_{i}}$ is constant (an analogous condition will be seen when the zone widths of Fresnel lenses are discussed in section \nameref{sect:fresnel}).

\subsection{Crystal diffraction}
\label{CrystalDiff}
\subsubsection{Ideal and mosaic crystals}
Before going into details of the calculation of diffraction efficiencies it is useful 
to introduce a distinction between `ideal' and `mosaic' crystals. In ideal (defect free)
crystals the
crystalline pattern is continuous over macroscopic distances. Examples of such crystals
are the highly perfect Silicon and Germanium crystals now commercially available thanks to
their great commercial interest and the consequent intense development effort. 
`{\textcolor{green}{Mosaic crystals}}' on the
other hand are described by a very successful theoretical model introduced by 
{\textcolor{green}{Darwin}}
\cite{darwin1914,darwin1922} a 
century ago. 
Darwin  modelled imperfect  crystals  as composed of a large number of 
small `crystallite' blocks, individually having 
a {\textcolor{green}{perfect crystal}} structure 
but slightly misaligned one to another.
The spread of the deviations, $\omega$, of the orientation
 of a lattice plane in one block from the mean
 for the entire mosaic crystal
is described by means of a probability distribution $\Omega '(\omega)$,
 the so-called mosaic distribution function. 

\begin{figure}[ht]
  \center
  \includegraphics[scale=0.7, keepaspectratio=true]{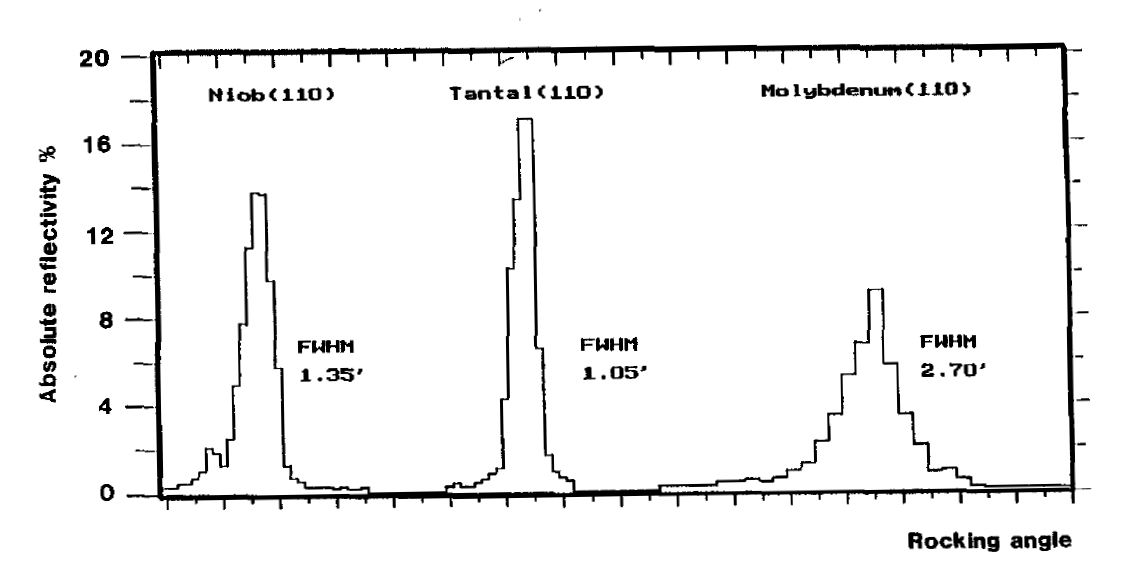}
  \caption{Measured rocking curves (at 412 keV) for a few metal crystals having 
  different degrees of mosaicity. The small number triplets shown along with the 
  element names are the 'Miller indices' (see \cite{zachariasen}) identifying the 
  lattice planes used for diffraction. Reprinted from \cite{lund1992}.}
  \label{fig:RockingCurves}
\end{figure}

Darwin's model has been very useful and quantitatively describes many aspects of real crystals.
The mosaic distribution function, $\Omega '$ can be found experimentally for a given crystal
through observation of the `{\textcolor{green}{rocking curve}}', which is the measured reflectivity as function of 
angle for an incident beam of parallel, monochromatic gamma-rays when a 
crystal is scanned through the angle corresponding to a Bragg-reflection. The crystal sample should be thin enough that the
reflectivity is never close to the saturation value.
Examples of measured rocking curves are shown in Figure~\ref{fig:RockingCurves}. 

For good quality crystals the mosaic distribution can be  well approximated by a Gaussian function.
Its width, as observed through the rocking curve,  is called the 
{\textcolor{green}{mosaic width}} of the crystal.
Mosaic widths are specified as angular quantities,
characterized by either the Full Width at Half Maximum (FWHM) or the standard deviation, $\sigma_\theta$, of the rocking curve. 

Rocking curves are measured at constant diffraction angle, hence constant energy.
For Laue lens design where
the source direction is the fixed quantity, what is often of interest is the mosaic distribution 
as a function of the energy offset from the energy, $E_B$, 
corresponding to the {\textcolor{green}{Bragg angle}}. If the standard deviation of the distribution as a function of energy is $\sigma_E$, then the two quantities are related by:
\begin{equation}
    \frac{\sigma_E}{E_B} = \frac{\sigma_\theta }{\theta_B}
 \label{dTheta-dE-relation}
\end{equation}
To be useful in a Laue telescope the rocking curve for each crystal should 
possess a single, narrow peak.
Unless great care is taken in the growth of crystals the mosaic distribution may not be well behaved and the rocking curves may be broad or exhibit multi-peaked structures.

\subsubsection{Diffraction efficiency}
 \label{diff_eff}
According to Schneider \cite{Schneider1981b}  the crystal reflectivity, $\nu (E)$, for mosaic crystals of macroscopic thickness in the
Laue-case can be calculated from:
\begin{equation}
   \nu (E) = \frac{1}{2} e^{-\mu (E) t} (1 - e^{-\Omega (E - E_0) R(E) t}).
 \label{eq:XtalRefl}
\end{equation}
Here $ \mu(E)$ is the linear attenuation coefficient for photons of energy $E$ and $t$ is 
the crystal thickness. $\Omega (E-E_0)$ is the mosaic distribution as
function of energy, and $R(E)$ is the specific reflectivity  (reflectivity per unit thickness). $E_0$ is defined as the energy
where $\Omega$ is at its maximum. 

In the following we shall assume that the mosaic distribution 
$\Omega$ has a Gaussian shape:
\begin{equation}
 \Omega (E-E_0) = \frac {1}{\sqrt{2\pi}\sigma} e^{-(E-E_0)^2 / {2 {\sigma^2}}} .
 \label{Omega}
\end{equation}
We then get for the peak reflectivity:
\begin{equation}
 \nu (E_0) = \frac{1}{2}e^{-\mu (E_0) t} (1 - e^{-\frac{R(E_0) t}{\sqrt{2\pi} \sigma}}).
 \label{PeakRefl}
\end{equation}
The value of the crystal thickness which maximizes the peak reflectivity is 
\begin{equation}
             t_{max} = \frac{\ln{(1+\alpha})}{\mu (E_0)\alpha}  
 \label{MaxThick}
\end{equation}
with:
\begin{equation}
             \alpha = \frac{R(E_0)}{\mu (E_0)} \frac{1}{\sqrt{2\pi} \sigma}
 \label{MaxThickAlpha}
\end{equation}
and the corresponding peak reflectivity is
\begin{equation}
              \nu_m(E_0) = \frac{1}{2} \alpha (1+\alpha )^{-\frac{1+\alpha }{\alpha }}.
 \label{MaxRefll}
\end{equation}
Note that the specific reflectivity, R, and the attenuation coefficient, $\mu$, are
characteristic  of a particular material and set of crystalline planes,
 whereas the mosaic width, $\sigma$, depends on the method of manufacture and subsequent
 treatment of the crystals. It is therefore reasonable to start by seeking crystals that
 maximise the value of $R / \mu$ and leave the choice of the mosaic width to the detailed
 lens design.
 
 The specific reflectivity is given by:
\begin{equation}
    R(E) = 2r_e^2 \frac{\lambda ^3}{\sin(2\theta _B)} \Big (\frac{F_{struct}(x)}{V}\Big )^2 e^{-2Bx^2}
 \label{SpecifRefl}
\end{equation}
 with $x = \sin (\theta_B) / \lambda$.
 Here $r_e$ is the classical electron radius, $V$ 
 is the volume of the unit cell and
$F_{struct}$ is the `structure factor' for the 
crystal unit cell. The structure factor 
depends on the crystal structure 
type (e.g. body centred cubic, face centred 
cubic), on the atoms, and on the choice 
of lattice planes involved, described 
by the Miller indices (h,k,l). The exponential 
factor describes the reduction in the diffraction 
intensity due to the thermal motion of the 
diffracting atoms. Considering only crystals 
of the pure elements, \ref{SpecifRefl} can be 
rewritten as:
\begin{equation}
  R(E) \propto E^{-2} a^{-5/3} (f_1(x,Z))^2 e^{-2B n^2 / {2d^2}},
 \label{SpecifElemRefl}
\end{equation}
where  $a$ is the atomic volume, 
$d$ is the interplanar distance of the diffracting planes, $f_1(x,Z)$ is the atomic 
form factor and $n$ is the diffraction order. Here use has been made of 
the approximations $\sin (2\theta_B) \sim 2 \sin (\theta_B)$ and $ x \sim n / 2d$. 

The linear attenuation coefficient, $\mu (E)$, can be expressed as %
a
function of the total atomic cross section, $\kappa (E)$, and the atomic volume :
\begin{equation}
\mu (E) = \frac{\kappa (E)}{a}
 \label{eq:LinearAtten}
\end{equation}
thus

\begin{equation}
 \frac{R}{\mu} \propto E^{-2}  a^{-2/3} \frac{f_1(x,Z)^2}{\kappa (E)} e^{{-B n^2}/{2d^2}} .
 \label{eq:R_mu_ratio}
\end{equation}

Since at high energies both $f_1(x,Z)$ and $\kappa (E)$ are roughly proportional 
to $Z$ it is clear that for gamma-ray energies a high atomic number and a small 
atomic volume (a high atomic density) are important for maximizing R/$\mu$.
For energies below ${\sim100}$ keV photoelectric absorption may rule out the use of crystals
of the heaviest elements for Laue lenses. 
\begin{figure}[!t]
  \center
  \includegraphics[scale=0.6,keepaspectratio=true]{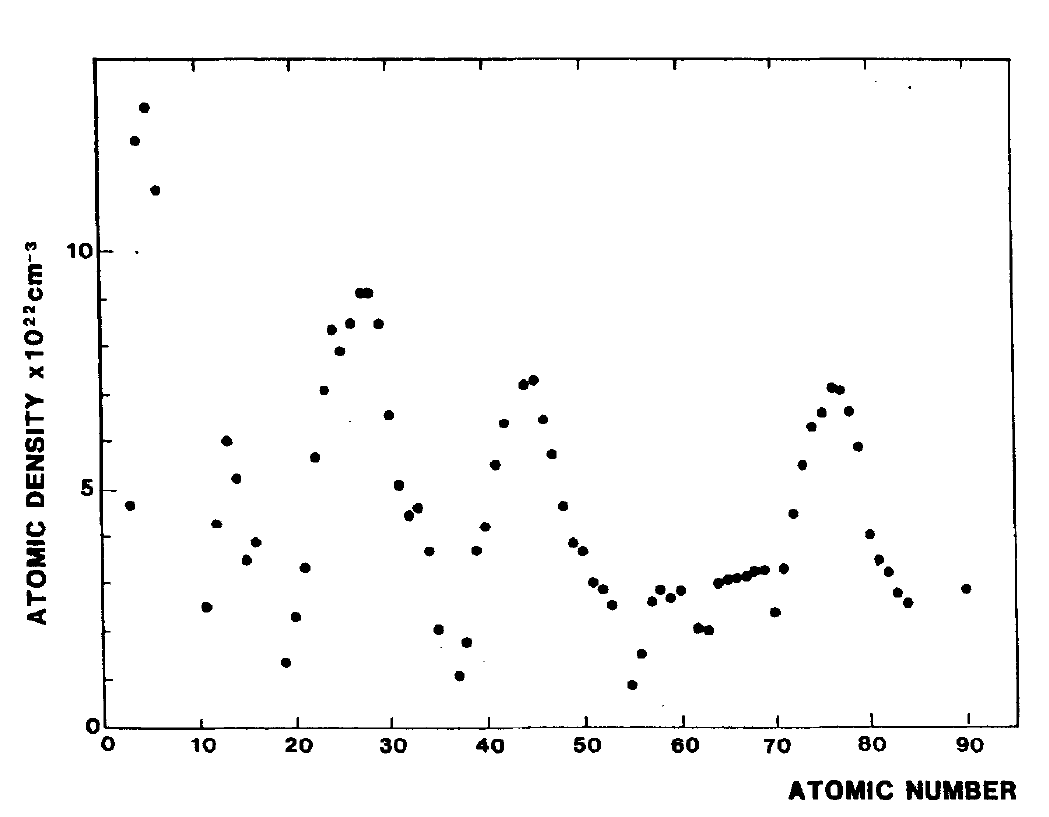}
  \caption{The Atomic Density as function of the Atomic Number, Z. Reprinted from \cite{lund1992}.}
  \label{fig:AtomicDensity}
\end{figure}

The atomic density of crystals of the pure elements varies systematically with the atomic number, $Z$, as illustrated in Figure \ref{fig:AtomicDensity}. 
The most suitable elements for Laue lens crystals are found near the peaks
in this plot, that is near $Z = 13$ (Al), $Z = 29$ (Ni, Cu,), $Z = 45$ (Mo, Ru, Rh Ag) or
$Z = 76$ (Ta, W, Os, Ir, Pt, Au). 

The atomic form factors are tabulated in the literature \cite{IntTblXCryst1977} and software
for their calculation is publicly 
available \cite{XOP-paper}.

The thermal factor, $B$, turns out to be anti-correlated with the atomic 
density, thereby strengthening somewhat the case for a high atomic density \cite{Warren1969}.

\subsubsection{Extinction effects}
\label{sect:Extinction}
In the above derivation of the diffraction efficiencies only the attenuation by 
incoherent scattering processes was explicitly considered in equation \ref{eq:LinearAtten}.
However losses due to coherent effects also occur.  These are termed 
`extinction effects' \cite{Chandrasekhar1960}. 
One type of extinction loss is due to the diffraction process itself and occurs in both
mosaic and perfect crystals. 
Photons are removed from both the incoming beam and from the diffracted beam by diffraction. 
(The diffracted beam is a mirror image of the incoming beam with respect to the lattice planes and fulfills the Bragg condition just as well). This dynamic interaction  is termed `secondary 
extinction' and accounts for the factor $\frac{1}{2}$ in \ref{eq:XtalRefl}. 

A more subtle extinction effect, which is only present if phase coherence is maintained 
through multiple diffractions, is termed the `primary extinction'. 
Every diffraction instance is associated with
a phase shift of $\pi /2$. Consequently, after two coherent diffraction processes the 
photon has accumulated a phase shift of $\pi$ and destructively interferes with the 
incoming beam. In the same way, three coherent diffraction processes will cause destructive
interference in the diffracted beam. This effect only occurs in perfect crystals or in mosaic crystals in which the size of the crystallites is large enough that there is a  
significant probability of multiple successive coherent
scatterings. The critical dimension here is the 
`extinction length' (see \cite{zachariasen}) which can be estimated as:
\begin{equation}
{
    t_{ext} \approx \frac{1}{r_0} \frac{V}{F_{struct}(x) \lambda}  \ }.
    \label{eq:extinct_length}
\end{equation}

The extinction length for the Cu(111) reflection at 412 keV is 66 $\mu$m  \cite{Schneider1981b}~. As $t_{ext}$ is proportional to the energy it will be at the 
lower gamma-ray energies that extinction effects may become noticeable.
For Laue lenses it is important to find or develop crystals in which the
defect density is high enough to keep the crystallite size below $t_{ext}$ at the 
lowest energies where the crystals are to be used. 

\subsection{Focusing elements}

\subsubsection{Classical perfect crystals}
\label{perfect_crystals}

Perfect crystals, where the ideal lattice extends over macroscopic dimensions, are not
particularly suitable for the use of Laue lenses in Astrophysics 
because they are too selective regarding 
the photon energy, even for Laue lenses intended for narrow line studies. 
Perfect crystals diffract with high efficiency, but only for an 
extremely narrow range of energy/angle combinations. For example,
at 511 keV a perfect
Germanium crystal will have an angular width  of the diffraction peak
(the `Darwin width') of only 
0.25~arc-seconds. This should be compared to the Bragg angle, which at this energy is 750~arc-seconds, {\emph{i.e.}} about one part in 3000.
The corresponding energy bandwidth is then only 0.14~keV! 
Thus, perfect crystals are not preferred for observations of astrophysical sources.

\subsubsection{Classical mosaic crystals}
\label{mosaic_crystals}

Fortunately, perfect crystals are not the norm. 
Most artificial crystals grow as mosaic crystals. 
According to the Darwin model, such crystals can be viewed 
as an ensembles 
of  perfect micro-crystals with some spread of their 
angular alignments. Photons of a specific energy may traverse 
hundreds of randomly oriented crystallites with little interaction 
and still be strongly diffracted by a single crystallite oriented
correctly for this energy. Mosaic crystals generally perform much 
better than perfect crystals in the context of Laue lenses. 
The internal disorder, the mosaic width, may be controlled to 
some extent during the crystal growth or by  subsequent treatment. 
Mosaic widths of some arc-minutes can be obtained with relative ease for 
a range of crystal types. For the lenses described below a 
mosaic width of about 0.5 arcminutes is typical. Such values can be obtained, 
but this has required substantial development effort \cite{courtois2005}. 
Copper crystals in particular have attracted interest because of the need for 
large size, high quality, Copper crystal for use in low energy neutron diffraction. 

\begin{figure}[!t]
  \center
 \includegraphics[scale=0.22,keepaspectratio=true]{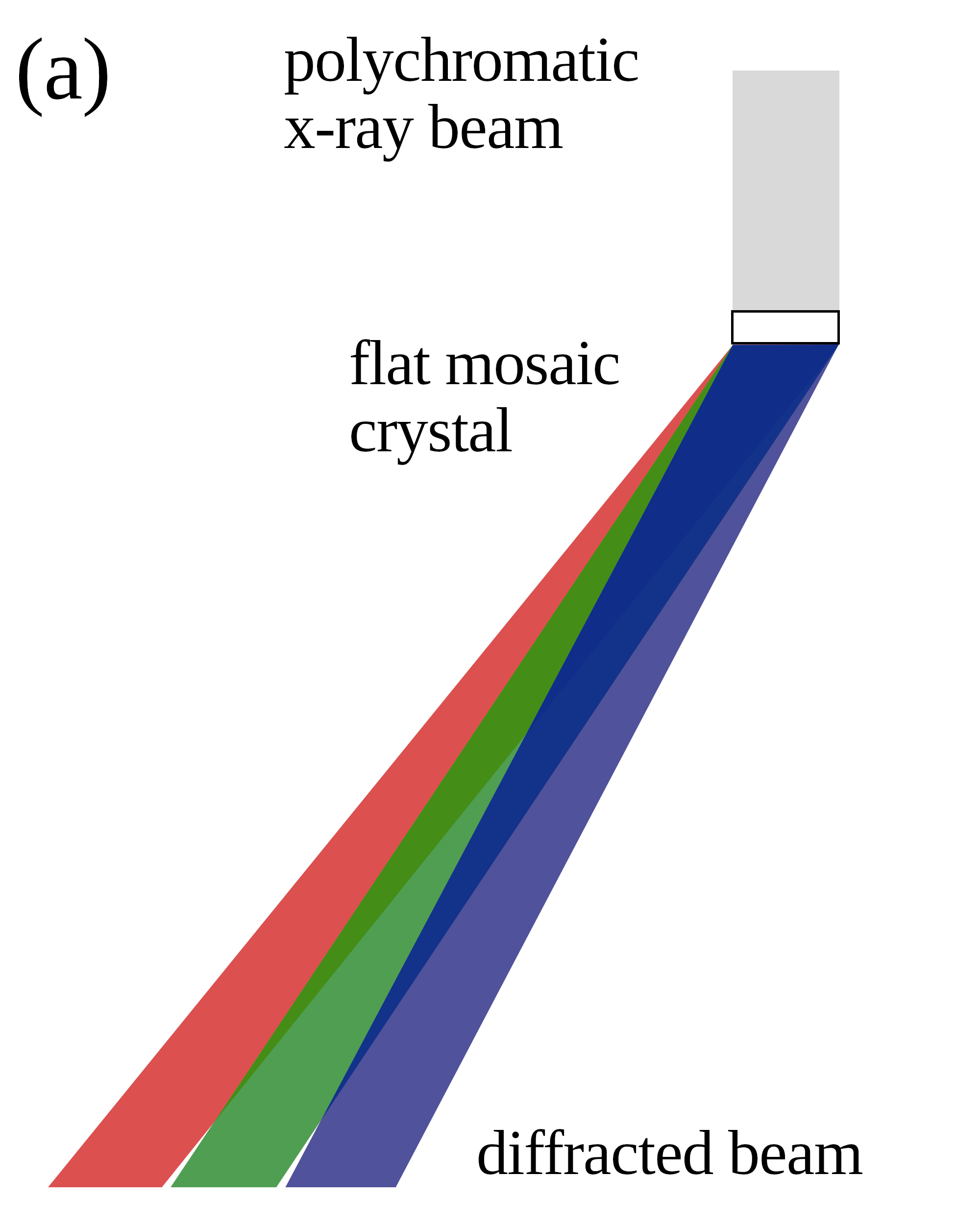}
 \includegraphics[scale=0.22,keepaspectratio=true]{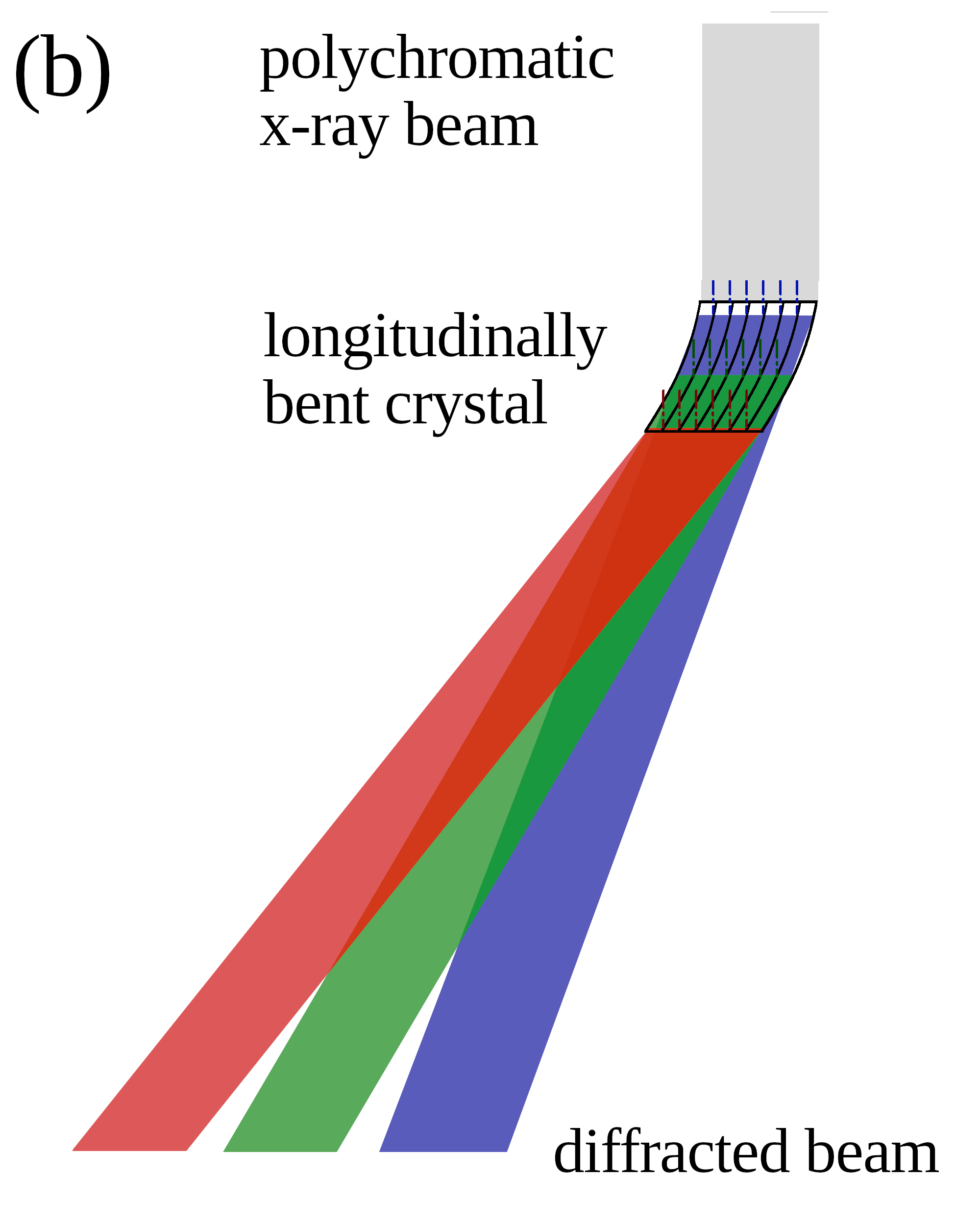}
 \includegraphics[scale=0.22,keepaspectratio=true]{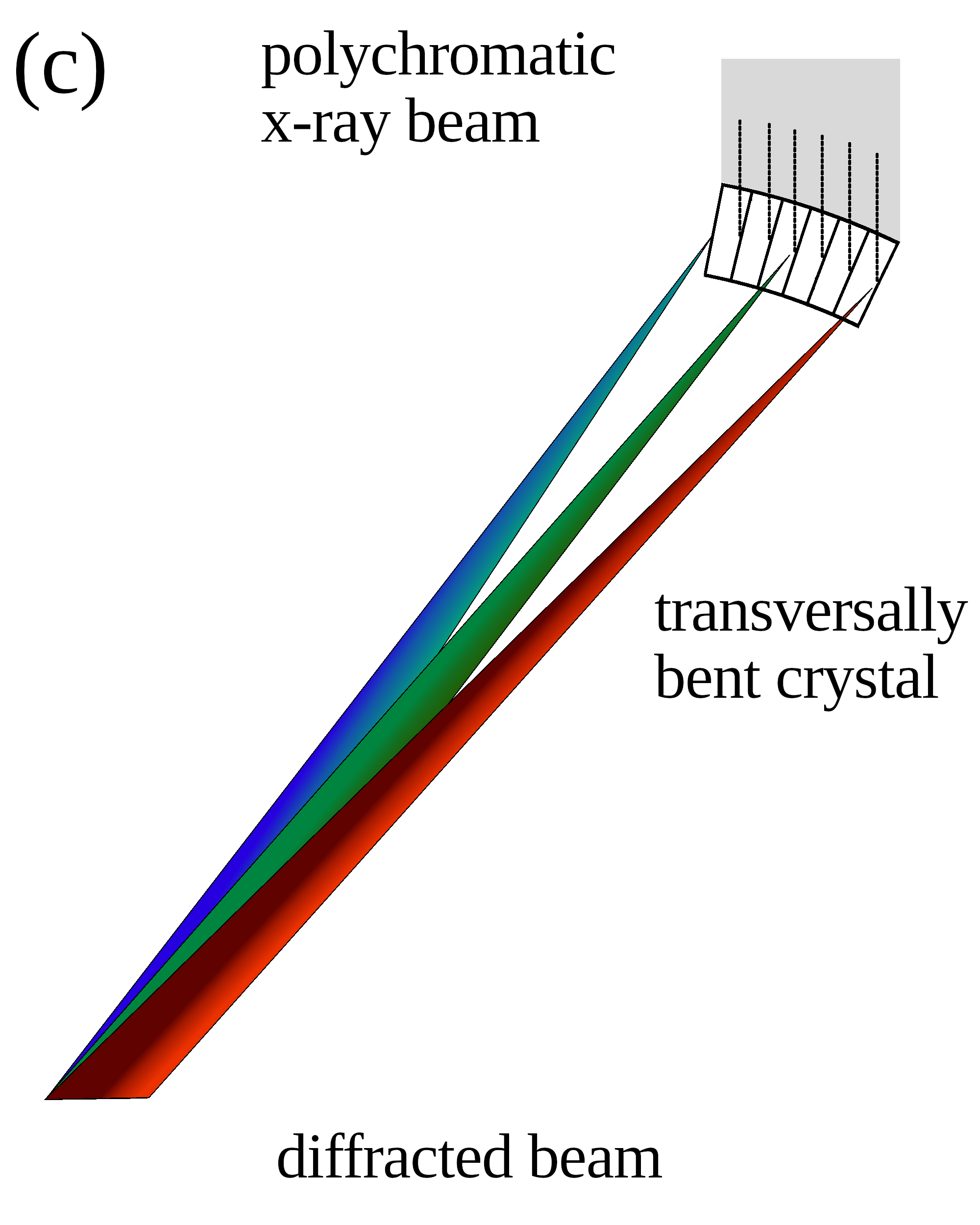}
\caption{Different crystal options proposed for 
Laue lenses. Flat mosaic crystals (a) produce 
a chromatic effect that is evident if the mosaicity  
is large compared with the angular size of the crystals. 
Longitudinally bent crystals (b)
can in principle reach higher diffraction efficiencies 
than mosaic crystals. However, they are complex to 
manufacture. A transverse bent perfect crystal (c)
can offer almost achromatic focusing at the expense 
of the effective area, as the geometric area
corresponding to a given energy is a fraction of 
the total surface of the crystal.
In this case the pass-band of a crystal depends 
on its curvature.
}
  \label{fig:LaueOption}
\end{figure}

It  must be kept in mind that Bragg's law is always strictly valid, even for 
mosaic crystals. As illustrated in Figure \ref{fig:LaueOption}-(a), after diffraction from 
a mosaic crystal a polychromatic beam of parallel gamma-rays with a spread of energies 
will emerge as a rainbow coloured fan. Its angular width will be twice the angular mosaic 
width of the crystal. Even if the crystal is oriented so that the central ray of the emerging beam hits the detector, the extreme rays of the fan may miss it. 

At a given energy, the radiation diffracted from a flat mosaic crystal forms in 
the detector plane a projected image of the crystal. It is important to note that 
this projected image does not move if the crystal tilt is changed. Its position 
is fixed by Bragg's law and it only changes in intensity.

\subsubsection{Crystals with {\textcolor{green}{curved lattice planes}}}

As well as perfect and mosaic crystals, a third group has 
attracted interest as potential diffractive elements for Laue lenses. These are  
perfect crystals with curved lattice planes. The curvature of the lattice planes has two
remarkable effects: i) the secondary {\textcolor{green}{extinction}} may be suppressed and  ii) the energy 
pass-band is not constrained anymore to the Darwin width but will be 
defined by the total range of lattice direction, {\emph{i.e.}} in principle something 
under our control. If the lattice curvature is correctly chosen relative to the photon
energy secondary extinction may be suppressed and the diffraction efficiency can approach
100 \% - ignoring incoherent absorption, which is always present.

Different methods to create such lattice curvature have been proposed and experimentally demonstrated. One involves  imposing a thermal gradient on the crystal along its thickness, such that the hot side expands and the cold side contracts. This method is very convenient in the laboratory because both the degree of bending (the bending radius) and the `sign' of the bending can be changed with minimal change in the experimental set up \cite{smither05b}. Unfortunately, this method cannot be used in space due to the significant power dissipation required to maintain the thermal gradient.

A second method relies on specific pairs of elements (or compounds) which can form perfect crystals across a range of component fractions. Stable, curved lattice planes exist in these binary crystals in the regions where a composition gradient is present \cite{Abrosimov05}. Silicon/Germanium composition-gradient 
crystals have been proposed for the `MAX' Laue lens described in Section \nameref{MAXproject}.

A further bending method relies on externally applied mechanical forces to bend the crystals.
Mechanically bent crystals are used in several applications in laboratory experiments
including monochromators. However the mass of the structures necessary to maintain the bending forces is unlikely to be acceptable for
a space experiment.

Controlled permanent bending of 
Silicon and Germanium wafers by  surface scratching has been 
developed by the Institute of Materials for Electronics and 
Magnetism, (IMEM-CNR) in Parma ~\cite{Buffagni11} in connection 
with the Italian {\it{Laue}} project~\cite{virgilli2014}. 
The lapping procedure 
introduces defects in a superficial layer 
of a few microns, providing a high compressive
stress resulting in a convex surface on the worked 
side. In such \textit{transversally bent crystals} the orientation of the diffraction planes  with respect to the incident radiation continuously changes in the direction of the curvature of the crystal. 
If the bending radius is 
equal to twice the focal length of the lens, the effect is to produce 
achromatic focusing as illustrated in Figure
\ref{fig:LaueOption}.

A spectrum of finite extent may be 
focused into an area which is considerably smaller than 
the crystal cross-section. In this way the overall  
Point Source Response Function (PSF) of a Laue lens can 
be narrower than achievable with flat crystals of the same size.

Transversally bent crystals were studied in
the {\it{Laue}} project in which a number of different 
aspects of the Laue lens technology were faced, from 
the production of suitable crystals to the definition 
of an accurate and fast method for their alignment. 
It was demonstrated through simulations and experimental tests 
\cite{virgilli13} that a transversally bent crystal
focuses a fraction of the radiation arriving on its 
surface into an area which, depending on mosaicity, 
can be smaller than the cross section of the crystal itself. 
For some crystals and crystallographic orientations, 
if an external (primary) curvature is imposed through 
external forces a secondary curvature may arise. This effect is a 
result of crystalline anisotropy. It has been termed 
quasi-mosaicity ~\cite{ivanov05} and leads to an 
increased diffraction efficiency and angular acceptance 
\cite{camattari11}.

\begin{figure}[ht]
  \centering
    \includegraphics[scale=0.18,keepaspectratio=true]{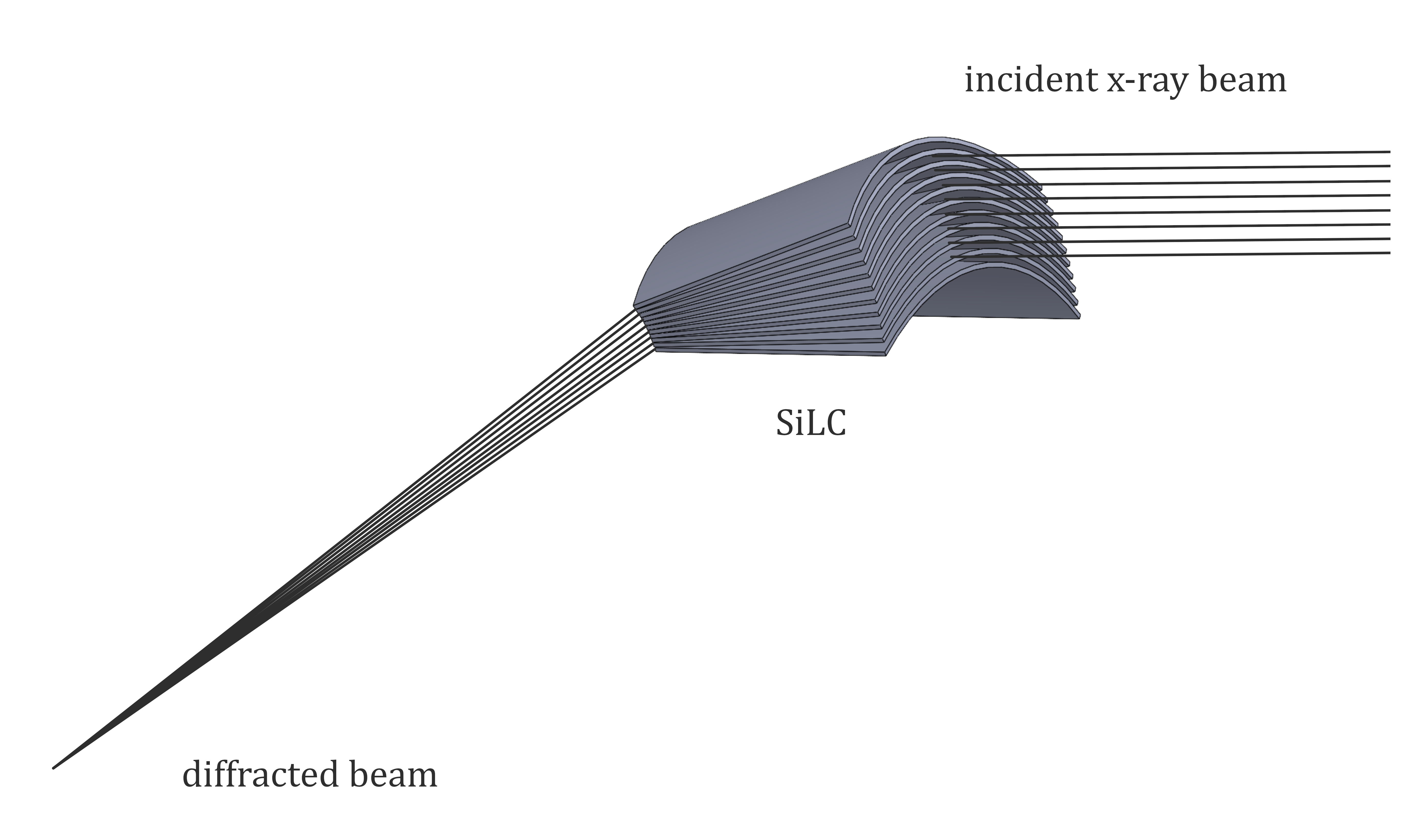}
  \caption{Sketch of the principle at the basis of the 
  SILCs elements developed at Cosine Measurement System (The Netherlands). 
  A polychromatic beam is focussed in two directions. 
Published with permission of the authors and adapted from Cosine 
Measurement Systems~\cite{cosine2022b}.}
  \label{fig:silcs}
\end{figure}

A promising manufacturing technology for bent crystals 
which may overcome some of the limitations of flat crystal 
diffraction optics is based on so-called Silicon Pore Optics (SPO)~\cite{bavdaz2012}. This is a bonding technology for silicon
wafers which is being developed for the ESA  {\athena} mission. 
It has made possible the development of novel units for focusing 
gamma-rays called Silicon Laue Components (SiLCs~\cite{ackermann13, girou17}). These components are being developed at the 
{\sc Cosine} company (The Netherlands) in collaboration with the University of 
California at Berkeley. They are self-standing Silicon diffracting elements which 
can focus in both the radial and the azimuthal directions.
SiLCs consist of a stack of thin Silicon wafers with a small wedge angle 
between adjacent plates such that the diffracted rays from all the plates converge at a 
common focus. 
The incidence angle of the radiation is 
small so the radiation passes through only one  plate.
The wafer angle with respect to the optical axis of the telescope is selected such that 
the mean angle enables diffraction at energy $E$, and the overall range of wedge angles between 
the wafers dictates the energy bandpass around the centroid $E$. 
As can be seen in Figure~\ref{fig:silcs} the curvature of the wafers allows focusing in the ortho-radial direction.


%
\subsection{Laue lens optimization}

The response of a Laue lens is strongly energy dependent. For a given focal length and crystal plane spacing 
the area diffracting a particular energy pass-band is inversely 
proportional to E. Furthermore, the diffraction efficiency of the crystals 
adopted for realizing these optics decreases with 
energy. These two reasons, combined with the fact that the gamma-ray emission of astrophysical sources typically decreases with energy according to a power law, make observations
at high energies even more challenging.
The dependence of the effective area on energy is a geometric effect and can be mitigated only at particular energies in narrow pass-band Laue lenses. The decrease in diffraction efficiency with energy can be mitigated by choosing crystals to  maximize the reflectivity.

A number of  parameters can be tuned in the optimization of a Laue lens. 
They are mainly related to the crystals properties (mosaicity, 
crystal material, diffraction planes, crystallite size, 
crystal thickness), or to the  overall lens structure 
(lens diameter, focal length, inner and 
outer radius, geometrical configuration). 
The optimization is complex and depends on the Laue
lens requirements (lens bandwidth, point spread function
extension, total weight of the lens). The main factors involved in the optimization are described in the following sections.

\subsubsection{Crystal selection}

As discussed in Section ~\nameref{diff_eff}, crystals with high 
atomic density and high atomic number are generally preferable as 
Laue lens elements except at the lowest energies 
where photo-electric absorption may render their use less attractive.
The technical difficulties involved in the fabrication and handling 
of crystals of the different chemical elements are also important 
factors. These difficulties vary significantly among the elements. 
The mechanical properties of the crystals are an important issue 
- for example  Silicon and Germanium are quite hard and rugged 
whereas Copper, Silver and Gold crystals are soft and 
require special care in treatment and handling.

As already observed in Sect.~\nameref{sect:Extinction}, the crystallite 
thickness plays an important role in the reflectivity optimization. 
For given values of the mosaicity and crystal thickness, 
the highest reflectivity is obtained for a crystallite size 
much smaller than the extinction length of the radiation.
At the energies of interest this thickness must be of the order 
of few $\mu$m. The crystal mosaicity also has a primary role 
in the Laue lens optimization. The higher the mosaicity,
the larger is the integrated reflectivity, and thus the effective area, but 
the broader is the signal on the focal plane detector.
These two effects act in opposite senses in the optimization of the 
sensitivity.

The crystal thickness is also an important factor for the optimization 
of the crystal reflectivity and therefore for the maximization of 
the Laue lens sensitivity. Eq.~\ref{MaxThick} provides the thickness 
that maximizes the reflectivity for a given material and for fixed 
diffraction planes. As the best thickness is also a function of energy
it is expected that, depending on the adopted geometry, crystals 
dedicated to high energy are thicker than those used to diffract
low energies. It must be also taken into account that, the choice 
of the thickness maximizing the reflectivity would often lead to
a mass unacceptable for a satellite-borne experiment. A 
trade-off between lens throughput and mass is then necessary.

\subsubsection{Narrow- and broad-band Laue lenses}
Depending on the scientific goal to be tackled, Laue lens can be designed, or adjusted, with two different optimizations: lenses for a broad energy pass-band (e.g. 100~keV - 1~MeV) or those configured to achieve
a high sensitivity over one or more limited range(s) of energy. The latter can be valuable for studying gamma-ray lines, or narrow band  radiation. Relevant energies of interest might be the 511~keV  e+/e- annihilation energy or the 
800-900 keV energy range for its importance in Supernova emission.
The two classes of Laue lenses need different optimizations and 
dispositions of the crystals over the available geometric area.  
\\
For a \textbf{narrow energy pass-band} Laue lens as many as possible of the crystals should be tuned to the same energy. According 
to Eq.~\ref{e_r}, the d-spacing 
of the crystals should ideally increase in proportion to their radial distance from 
the focal axis in order keep  the diffracted energy fixed. 
\\
The energy range of a \textbf{broadband Laue lens} follows from Eq. 
 \ref{e_r}. With the focal length and d-spacing of the 
 crystalline diffraction planes both fixed, the energy range will be from $$E_{min} = \frac{hc~F}{d_{hkl}~R_{max}} \: \: \textrm{ to } \: \:
 E_{max} = \frac{hc~F}{d_{hkl}~R_{min}},$$
where the  
 radial extent of the lens is R$_{min}$ to R$_{max}$.
If the inner and outer radii are fixed, the simultaneous use of different 
materials, and thus of different d$_{hkl}$, would allow enlarging the 
Laue lens energy pass-band compared with a single-material Laue lens.
Equivalently, for given energy pass-band and focal length, the use of multi-material  crystals would allow a more  compact lens.

\subsubsection{Tunable Laue lens}
\label{tunable}
\label{sect:tunable}

\begin{figure}[ht]
\begin{center}
\includegraphics[scale=1.1]{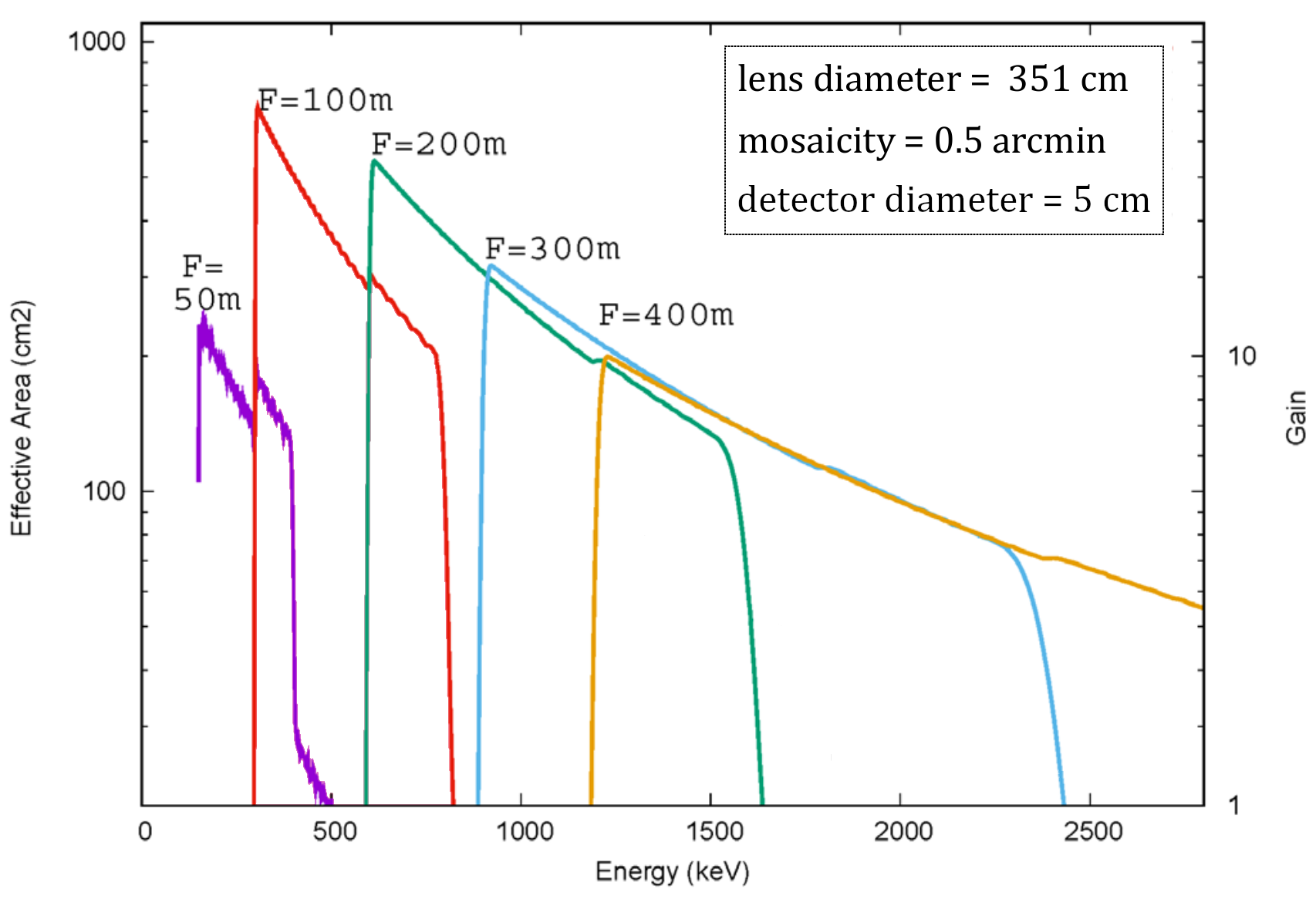}
\caption{The effective area of a single Laue lens which is tunable for
different energy passbands. The adjustment of the lens to a different band involves changing
both the focal length and the orientation of each crystal. When the focal length increases from 50 to 400~m,
the integrated effective area increases and the pass-band becomes larger. Note that the detector size assumed throughout is matched to the 100~m focal length configuration.  Figure 
reprinted with permission and adapted from\cite{Lund2020}.}
\label{fig:tunable}
\end{center}
\end{figure}

A classical Laue lens with fixed inner/outer radius and focal length
has a pass-band which is uniquely defined by the d-spacing 
of the crystals used. The red curve in Figure ~\ref{fig:tunable} shows the 
effective area of an example  100~m
focal length  Laue lens configured to cover a 300 - 800~keV energy pass-band.

If all of the crystals could be retuned for different focal lengths, the Laue less could be made sensitive to different pass-bands. Furthermore, as shown in
Figure ~\ref{fig:tunable}, the larger the focal length, the broader the pass-band and the higher  the integrated effective area. The adjustment in orbit is not trivial - both  the orientation of thousands of crystals and the lens to detector separation must be changed and verified. The former requires thousands of actuators and a sophisticated optical system. 

An innovative mechanism for the adjustment of the orientation of a crystal, along with an optical system for monitoring alignment of each one, has nevertheless been proposed~\cite{lund2021a, lund2021b}.
The mechanism is based on a miniature piezo-actuator 
coupled with a tilt pedestal and does not require power 
once a crystal has been correctly 
oriented. It is assumed that the lens and detector are 
on separate spacecraft that can be manoeuvred to adjust 
their separation. 

\subsubsection{Multiple layer Laue lenses}
\label{sect:multiplelayer}

A possible way to increase the flux collection from 
a lens is to use two or more layers of crystals covering
the same area. For instance, two layers of crystals can 
be used, one on each side of the lens structure.
In order to focus at the same position, crystals 
placed at the same radius but in different layers 
must diffract at the same angle, so different crystals or 
Bragg planes must be used to  diffract different energies. 

\begin{figure}[h]
  \center
   \includegraphics[scale=0.36,keepaspectratio=true]{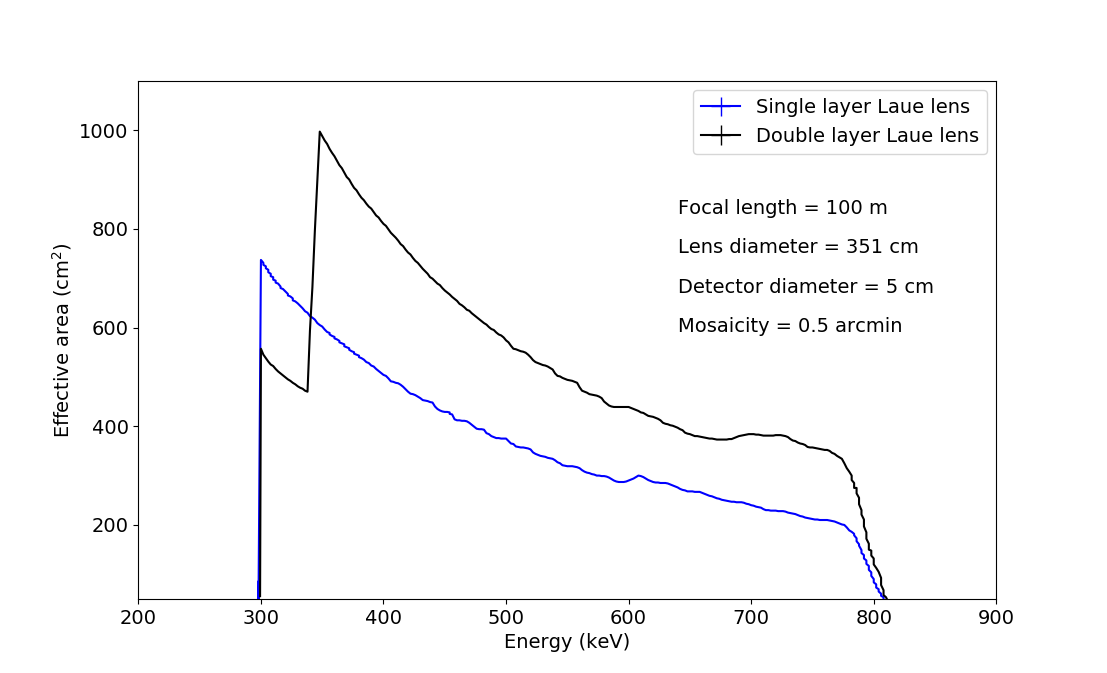}
  \caption{Blue line: effective area of a single layer of Ag(111) 
  crystals with optimized thickness. Black line: effective area of a Laue lens 
  made with two layers of crystals Ag(111) and Ag(200) whose thickness has been reduced with respect to the optimized thickness in order to save mass. 
  In spite of the thickness reduction, the effective area increases by about 65\%  and the overall mass is reduced by $\sim$10\%. Both effective areas are obtained with 100~m long focal length, crystals with 0.5~arcmin mosaicity and detector diameter of 5~cm. Figure taken and adapted with permission from \cite{lund2021a}.}
  \label{fig:multilayerLaue}
\end{figure}

In a simulations \cite{lund2021a} 
two layers of thin crystals made with Ag(111) 
and Ag(200), increased the 
effective area by about 65\% (see Fig. 
\ref{fig:multilayerLaue}). A third layer did not further 
increase the lens throughput. It must be stressed that 
with multiple layers, the diffracted 
radiation from any one layer will be 
attenuated by all of the other layers.
The number of layers maximizing the 
effective area will depend on the crystals parameters 
(thickness, mosaicity, diffraction efficiency).


\subsubsection{Flux concentration and imaging properties of Laue lenses}
\label{PSF}

The sensitivity of a telescope using a Laue lens depends on the effective area over which flux is collected, but because observations will almost always be background limited  it is also a function of the extent to which the collected flux is  concentrated into a compact region in the detector plane.

For a given lens design the collecting area at a particular energy is just the sum of the areas of the crystals multiplied by their reflecting efficiency at that energy. Obviously  only those crystals for which the incidence angle is close to the Bragg angle need be considered. In practice this means that the  only crystals that contribute are those with centres that fall inside an annulus whose width depends on the extent, $\Delta\theta$, of the rocking curve.  
For broadband lenses the incidence angle is a simple  inverse function of radius from the axis. Consequently crystals at the centre of the band will contribute most, with the response decreasing towards the edges of the annulus. Making the small angle approximation $\theta = r / 2F$,  the width of the annulus is given by:

\begin{eqnarray}
\Delta r = \frac{\Delta\theta}{d\theta/dr} = 2F \Delta \theta .
\end{eqnarray}

As the radius of the annulus is also proportional to $F$ this means that its area is proportional to $F^2$.

Because each crystal simply changes the direction of the  incoming parallel beam it will illuminate  a region in the detector plane identical in size and shape to the profile it presents to the incoming radiation. This immediately means that the focal spot can never be smaller than the size of the crystals.
Moreover, crystals that are not at the centre of the illuminating annulus will have a `footprint' in the detector plane that is offset from the instrument axis by a corresponding distance. Thus to a good approximation the radial form of the on-axis PSF of a broadband Laue lens is the convolution of the (scaled) rocking curve and a flat-topped function corresponding to the crystal size\footnote{The azimuthal extent of the crystals has been ignored and it is assumed that the crystals are smoothly distributed in radius.}.
 
 \begin{figure}[!t]
  \center
\includegraphics[scale=0.25,keepaspectratio=true]{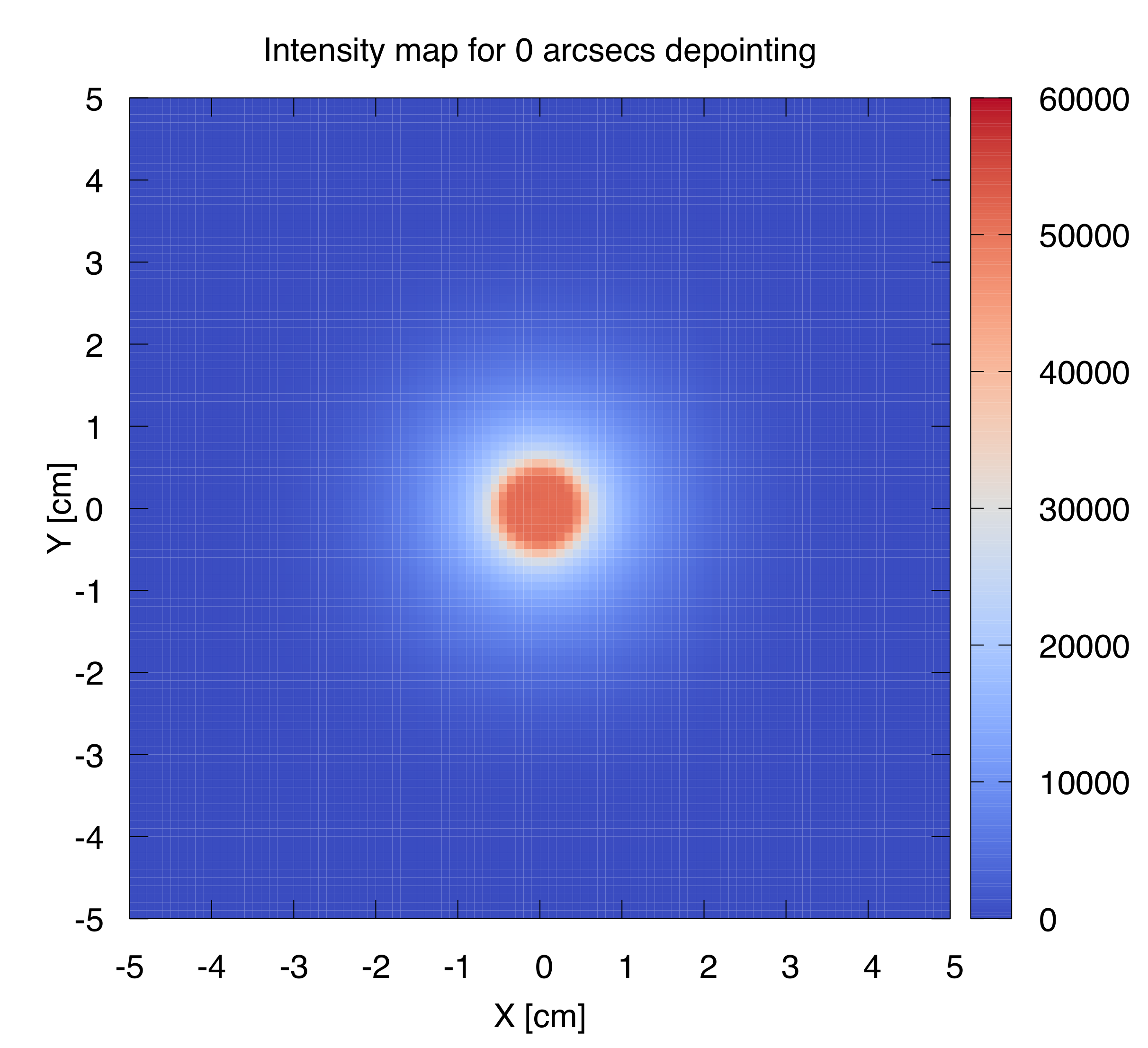}  \includegraphics[scale=0.25,keepaspectratio=true]{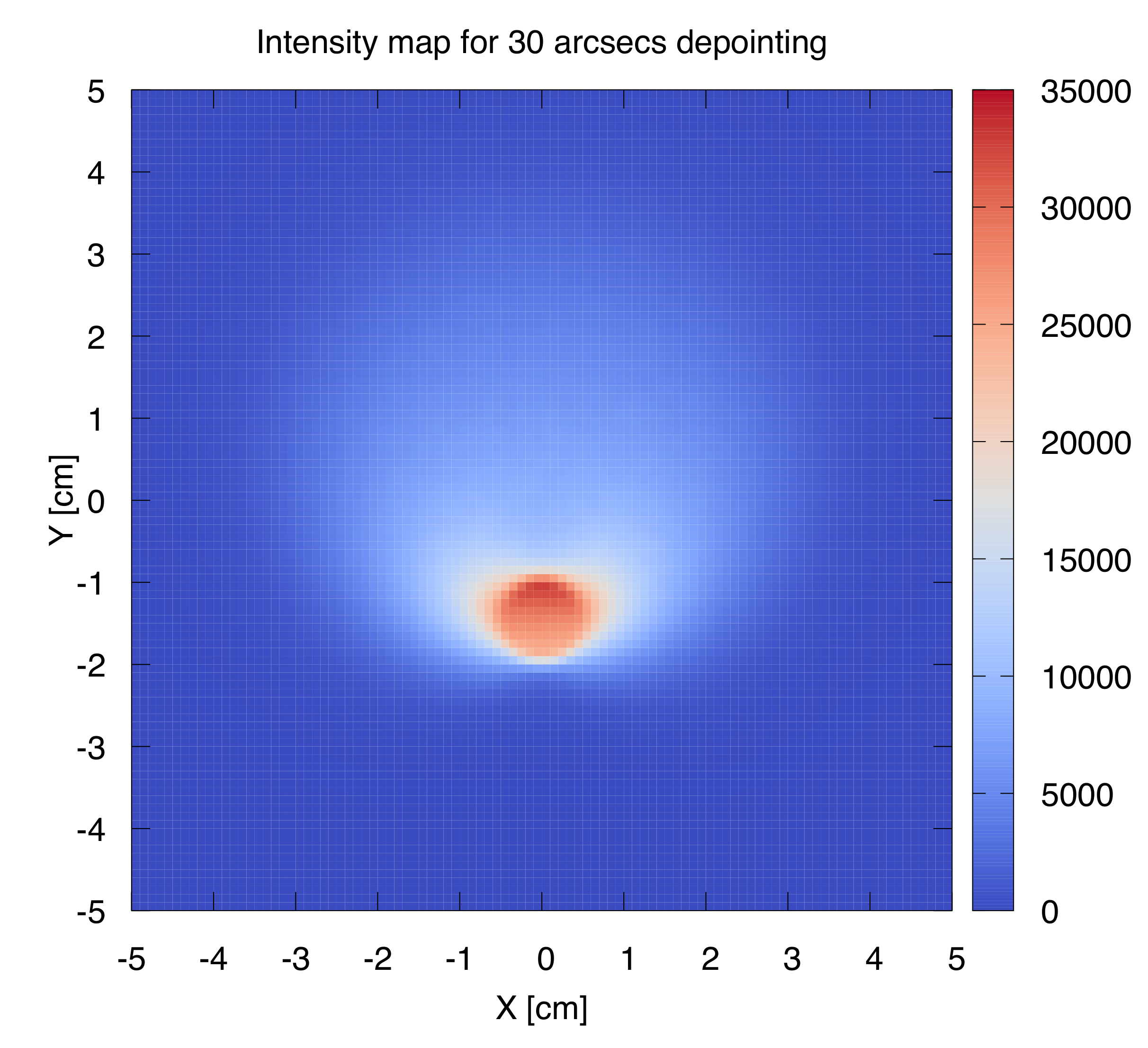}
\includegraphics[scale=0.25,keepaspectratio=true]{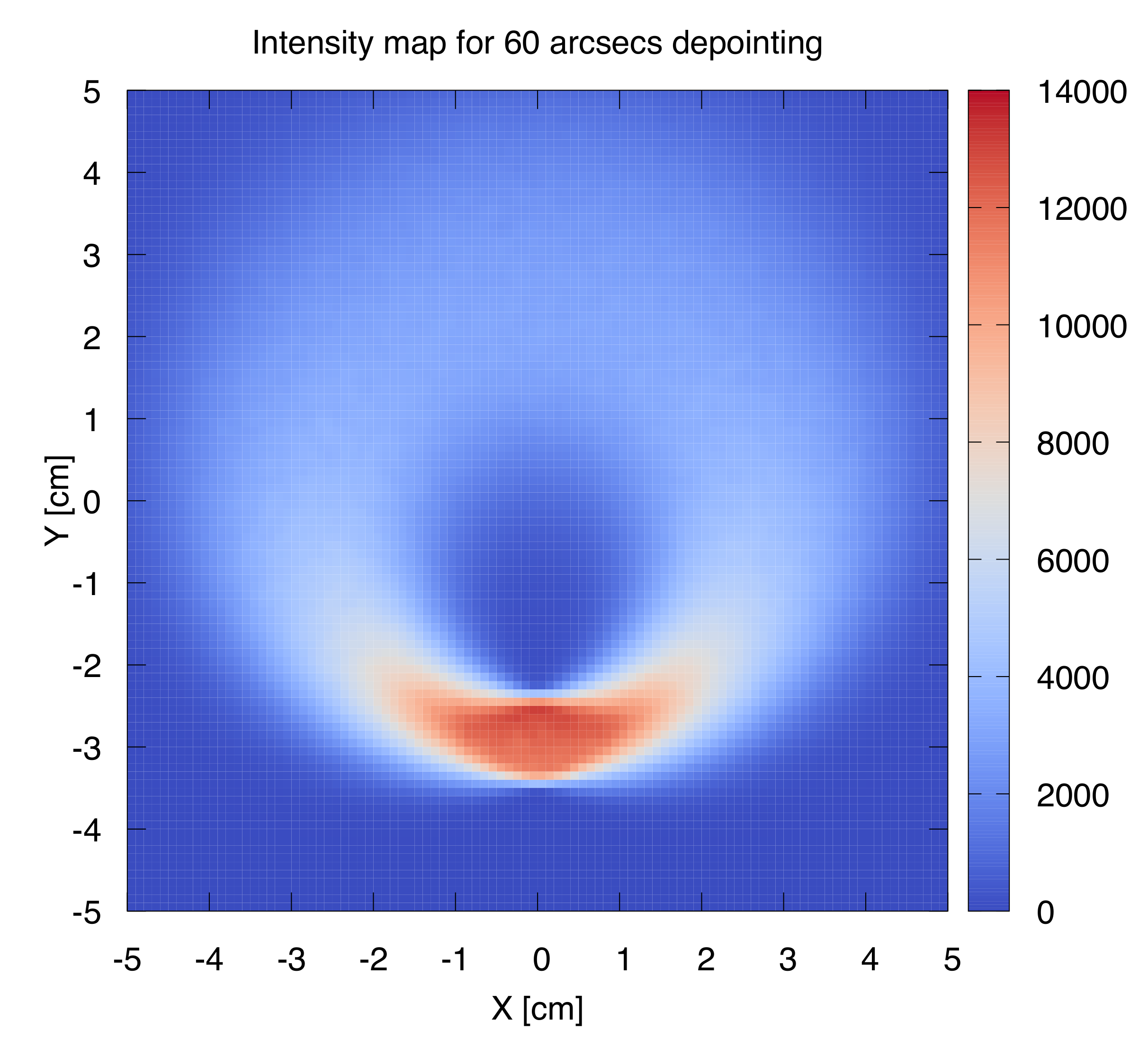}
  \caption{On- and off-axis images of a point source calculated for the lens of the
  Gamma Ray Imager proposal (see Sect.~\nameref{gri}), based on Copper crystals 
  and 100~m focal length. The image aberrations are typical for single-reflection devices.}
  \label{fig:ImageAberration}
\end{figure} 
As is usual with telescopes, other things being equal, the best signal to noise ratio will be obtained with a compact PSF. 
\begin{itemize}
\item{ In circumstances where use of smaller crystals is feasible and would lead to a significantly narrower PSF then choosing them will always offer an advantage.}
\item{ If the PSF width is dominated by the rocking curve width the situation is more complex. Decreasing the mosaicity will then narrow the PSF but also decrease the area of the diffracting annulus, reducing the advantage to be gained. }
\item{ Lastly because  of the $F^2$  dependence   of the area of lens over which crystals diffract a given energy, a wider PSF can actually provide an advantage if it  is the result of an increased focal length - although the signal will be spread over a detector area proportional to $F^2$, the uncertainty due to background only increases as $F$. This of course assumes that such a  design is feasible given the cost and mass of the larger lens and detector.}
\end{itemize}

The above discussion concerns the response to an on-axis
source. As Laue lenses are single reflection optics, they 
do not fulfill the Abbe sine condition and therefore 
for off-axis sources they are subject to coma. 
The aberrations  
are very severe as illustrated in
Figure ~\ref{fig:ImageAberration}.
The integrated flux is little affected by off-axis 
pointing, but importantly the image is smeared over a 
larger area and thus the signal-to-noise ratio of the 
signal is significantly reduced.


\subsection{Technological challenges}
\label{challenges}

 The development 
 of Laue lenses presents several technological 
 challenges. These can be divided into two categories.
The first is associated with the search for, and  production of, suitable materials and components.  
Highly reproducible crystals with high reflectivity are needed, as are 
thin, rigid and low Z substrates and structures to minimize the absorption.  The second category is related to the required accuracy of positioning and alignment, both for the  mounting of the crystals in the Laue lens and for the alignment of the lens with respect to the focal 
plane detector. Some of the main issues that are being faced in studies and development of Laue lenses are described below.

\subsubsection{I. Production of proper crystals and substrate} 
\label{crystals}

In order to cover a competitive geometric area  a Laue lens must contain a large number of crystal tiles.
The production of a large quantities of crystals having the optimal properties for providing a
reflectivity that is as high as possible is still problematic. Crystal growth has been described as an art as well as a science.
Advanced technologies present their own problems. For instance, it has been shown that bent crystals reduce the width of the PSF compared with 
flat crystals, but the advantages depend on very accurate  
control of the curvature.

\subsubsection{II a. Crystal mounting methods and accuracy} 

In a Laue lens it obviously important that each crystal is properly oriented. It is useful to consider three angles describing the orientation of a generic crystal: 
\begin{itemize}
\item (i) a rotation $\alpha$ about an axis parallel to the instrument axis.  If there is an error in $\alpha$, diffraction will occur with the expected efficiency but the photons will arrive in the focal plane displaced by a distance $R \Delta\alpha$, where $R$ is the distance of the crystal from the instrument axis. This displacement should be kept small compared with the spatial scale of the PSF. \\
\item (ii) a rotation $\phi$ about an axis normal to the diffracting crystal planes. To first order an error in $\phi$ will not have any effect on efficiency or imaging.\\
\item (iii) a rotation $\theta$ about an axis in the plane of the crystal plane and orthogonal to the instrument axis.
If $\theta$ does not have the intended value then the energy at which the reflectivity is highest will change. For any given energy, the position in the focal plane where photons of a given energy are arrive  will not be altered. However the number of photons diffracted by the crystal may either increase or decrease depending on the energy considered.
\end{itemize}

The probable effect is a spreading of the PSF as a result of enhancing the response of crystals that are not at the optimum radius for diffracting photons of a given energy towards the centre of the focal spot, at the expense of that of those that are. To avoid such spreading, errors in $\theta$ should be kept much smaller than the rocking curve width.

The most obvious mounting method is to use adhesives to bond the crystals to a supporting structure 
at their proper position and orientation.
However due to glue deformation 
during the polymerization phases it has not proven easy
to maintain the necessary precision \cite{2014NIMPA.741...47B,2015SPIE.9603E..08V}.  The 
amount of misalignment that is introduced depends on 
the type of adhesive used and on the polymerization 
process (two components epoxy adhesive, UV curable, 
thermal polymerization, etc.).

In the {\claire} experiment  the effects of uncertainties in the bonding process were avoided by the use of a manual adjustment mechanism (Sect:~\nameref{sect:claire}), but this technique is probably not appropriate for space instrumentation.

\subsubsection{II b. Laue lens alignment}
Because the MeV sky is poorly known at present, the issue of how to verify the
correct alignment of gamma-ray diffraction instruments after launch and during operations is important. 

It is therefore suggested that some optical means of verifying the shape of the large-scale structure of the lens should be incorporated from the start of the project. One possible way is to install mirror reflectors on the lens structure, the orientation of which can be monitored, perhaps from a separate 
detector spacecraft. If the attachment technique for these mirrors is similar to
that used for the Laue crystals the long-term stability of the 
mirror alignment will also give some confidence in the stability of the crystal mounting.


%
\subsection{Examples of Laue lens projects}
In this section we will review 
the Laue lens experiments that have been
realized or proposed from the early 2000s 
until the present. The CLAIRE project is the only Laue lens instrument
that has actually flown.  
Other experiments have been realized in the  laboratory as R\&D 
projects and were directed to the advancement of well-defined aspects of the Laue lens technology (mainly crystal 
production and tiles alignment). Studies have been conducted of several possible space missions based on Laue  lenses. 

\subsubsection{The {\textcolor{green}{CLAIRE}} balloon project (2001)}
\label{sect:claire}
\begin{figure}[ht]
\begin{center}
\includegraphics[scale=0.35]{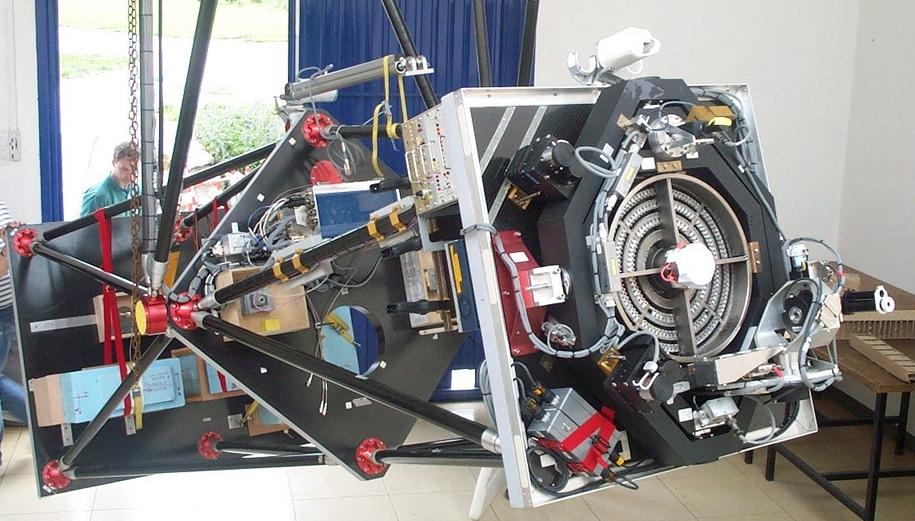}
\caption{The {\claire} telescope during preparation for the 2001 
balloon launch.  The instrument features, in addition to the 
Laue lens, a $3{\small\times}3$-array of cooled germanium 
detectors, and a pointing platform for the lens. The total weight 
of the payload was less than 500 kg. Published with permission from~\cite{vonballmoos04}.}
\label{claire:instrum}
\end{center}
\end{figure}

A pioneering proof-of-principle Laue telescope, CLAIRE, was 
built and successfully flown as a balloon payload by the 
Toulouse group in 2001 and 2002 \cite{laporte2000, halloin04}.
The balloon payload is shown in Figure~\ref{claire:instrum} 
during flight preparations.

\begin{figure}[ht]
  \center
  \includegraphics[scale=0.35,keepaspectratio=true]{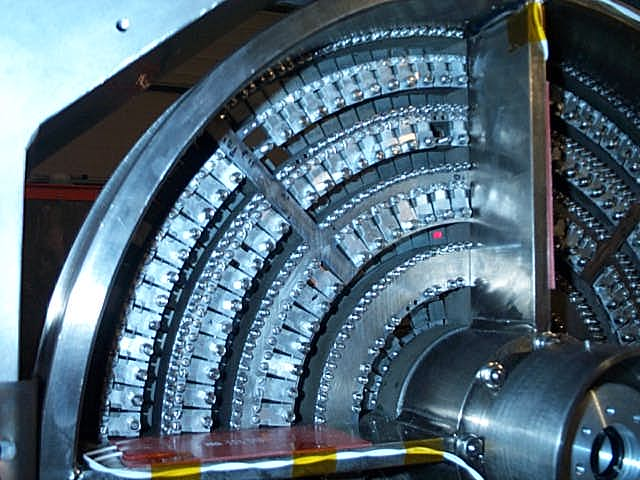}
  \caption{Detail of the {\claire} lens with 659 Ge/Si-crystals 
  mounted on adjustable lamellae on a Titanium frame.}
  \label{fig:claire:lens}
\end{figure}

The  {\claire} experiment is an example of a narrow energy pass-band instrument. 
It was designed for focusing photons with energy $\sim 170$~keV, with 
a pass-band of $\sim$ 4~keV at a focal distance of 2.8~m.
The lens (Figure~\ref{fig:claire:lens}) consisted of 659 Germanium/Silicon mosaic crystals arranged in 8 
concentric rings
providing a collecting area of 511 cm$^2$ and a field of view ({\sc fov}) of 90~arcseconds.
The mosaicity of the  crystals selected was in 
the range 60 to 120 arc-seconds, leading to an
angular resolution for the instrument of 25-30 arcseconds.
Two crystal sizes were used: 10 $\times$ 10 mm$^2$ and 7 $\times$10 mm$^2$. 
The crystals focused the radiation onto a 
Germanium detector with 9 elements, 
each 15 $\times$ 15 $\times$ 40~ mm$^3$, having
an  equivalent volume for background noise of 18 cm$^3$.

The fine tuning of the lens utilized a mechanical system capable of
tilting each crystal tile until the correct 
diffracted energy was detected.
The crystals were mounted via flexible lamellae on a rigid Titanium frame. 
The tuning of the individual crystals was done manually with adjustment screws.
Due to the finite distance (14~m) of the X-ray source during the crystal tuning 
phase, the gamma-ray energy used was lowered from 170 keV to 122~keV. 
Moreover, also due to the finite source distance used for the crystals alignment,
only a limited fraction of each crystal was 
effectively diffracting at the tuning energy, their subtended angle (about 150 arc-seconds, as seen from the source) being much larger that the crystal mosaicity. 

The performance of the complete lens was verified using a powerful industrial 
X-ray source at a distance of 200~m, (hence a diffracted energy of 165.4 keV). 
Seen from a distance of 200~m each crystal subtended an angle of about 10~arc-seconds, 
i.e. significantly less than the crystal mosaic width. 

The {\claire} lens was flown twice on balloon campaigns in 2000 and 2001. In both
flights the target source was the Crab Nebula. The observed diffracted signal at 
170~keV was found to be consistent with that expected from the Crab Nebula given the lens peak efficiency, estimated at about 8\% from 
ground measurements \cite{vonballmoos05}.

\subsubsection{The MAX project (2006)}
\label{MAXproject}
Following up on the successful {\claire} project the French Space Agency, CNES, 
embarked on the study of a  project, `MAX', for a satellite mission with a Laue lens. 
MAX was planned as a medium-sized, formation-flying project with separate lens and 
detector spacecrafts launched together by a single launcher. 

\begin{figure}[ht]
 \begin{center}
 \includegraphics[scale=0.5]{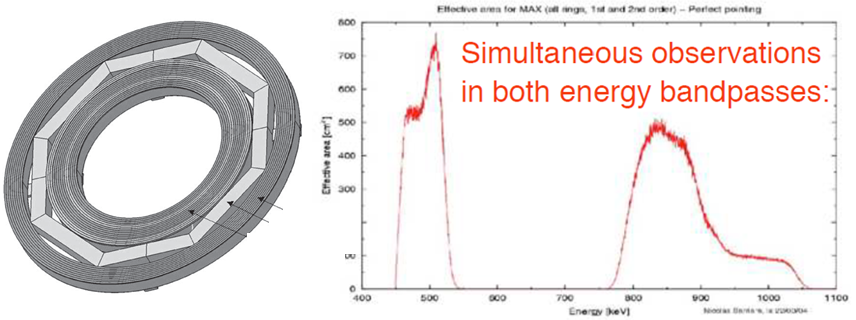}
   \caption{Left panel: The proposed MAX lens layout. Note the two crystal groups 
   corresponding to the two energy bands of the lens. A stable structural octagon
   supports both. 
   Right panel: The simulated response of the MAX Laue lens. More detailed analysis
   of the response using Monte Carlo techniques indicated that only about half of the 
   photons collected by the lens are sufficiently well focused to be of use during
   background limited observations. From \cite{Barriere2006}.
   }
 \label{fig:MAXlensLayout}
 \end{center}
\end{figure}

The scientific aims were (i) the study of supernovae type 1a (through observations 
of the gamma-lines at 812 and 847~keV),  (ii) a search for localized sources of  
electron-positron annihilation radiation at 511~keV and (iii) a search for 
478 keV line emission  from $^7$Be-decay associated with novae. 
These objectives could be met by a Laue lens with two energy 
passbands: 460-530~keV and 800-900~keV.  The left panel of Figure \ref{fig:MAXlensLayout} shows the lens proposed for MAX which would contain nearly 8000 crystal tiles 
15 $\times$ 15 mm$^2$ with a mosaic spread of 30 arc-seconds. The total mass of the 
Laue crystals was expected to be 115 kg. A focal length  of 86~m was foreseen. The predicted response of 
the MAX lens is shown in the right panel of Figure \ref{fig:MAXlensLayout}. 
In the end, however, CNES decided not to continue the development of MAX.

\subsubsection{GRI: the Gamma-Ray Imager (2007)}
\label{gri}

The Gamma Ray Imager (GRI) mission concept was developed by an international 
consortium and proposed to the European Space Agency in response to the 
`Cosmic Vision 2015–2025’ plan. 
GRI consisted of two coaligned telescopes each with a focal length of 100~m:
a hard X-ray multilayer telescope working from 20~keV up to 250~keV and a 
Laue lens with a broad passband  220~keV - 1.3~MeV.
The low energy limit of the GRI Laue lens was driven by the anticipated upper 
limit of the multilayer technology.  The NuSTAR mission has demonstrated the 
capability of multilayer telescopes of focusing photons up to 70-80 keV. 
Further developments are expected to allow multilayers to be used up to 200/300~keV.

The two optics were proposed to share a single solid state detector 
using  CZT (Cadmium Zinc Telluride) crystals which are attractive as they can be used  without 
cooling. Thanks to the 3-d capability of pixellated CZT, 
GRI could also be exploited for hard X-/soft gamma-ray polarimetry.
Due to the long focal length, a two spacecraft, formation flying mission 
was proposed. With these features, GRI was expected to achieve 30~arcsec 
angular resolution with a field of view of 5~arcmin. Unfortunately, the 
mission was not selected by CNES or ESA for further assessment.

\subsubsection{ASTENA: an Advanced Surveyor of Transient Events and
Nuclear Astrophysics (2019)}
\label{astena}

Within the European AHEAD project (integrated Activities in the High Energy 
Astrophysics Domain) \citep{natalucci18}, a mission was conceived to address some of 
the current issues in high energy astrophysics: a high sensitivity survey of transient
events and the exploitation of the potential of gamma-ray observations for  Nuclear Astrophysics.

\begin{figure}[ht]
  \center
  \includegraphics[scale=1.2,keepaspectratio=true]{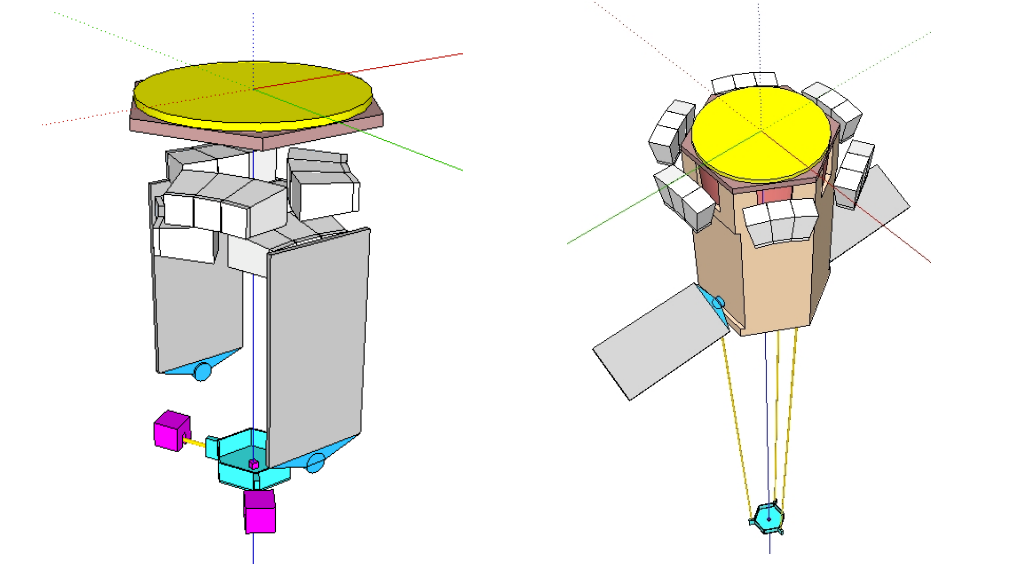}
  \caption{The ASTENA mission concept in its folded configuration 
  at launch (left) and in the operative configuration (right) in 
  which the WFM-IS array and the NFT focal plane detector 
  are unfolded.} 
\end{figure}

This mission concept, named ASTENA (Advanced Surveyor of Transient Events and Nuclear
Astrophysics) \citep{frontera19wp, guidorzi19wp} 
has been proposed to the European Space Agency in response to the
call "Voyage 2050". It consists of a Wide Field Monitor, 
with both spectroscopic and imaging capabilities (WFM-IS), and a Narrow Field Telescope (NFT) based on a broad energy pass-band (60 - 700 keV) Laue lens with a field of view of about 4~arcmin  and with an angular resolution of $\sim$30 arcsec. 
The Laue telescope exploits bent tiles of Si and Ge using the (111) reflecting 
planes. The tiles have dimensions of 30 $\times$ 10 mm$^2$ in size (the longer 
dimension being in
the focusing direction) and a bending curvature radius of 40~m. The focal length of the lens is 20~m. 
The crystals are arranged in 18 rings, with an outer diameter of $\sim$ 3~m 
that gives an outstanding geometric area of about 7 m$^2$. The focal plane 
detector consists of 4 layers of CZT drift strip detectors  \citep{kuvvetli05} 
each layer having a cross section of 80 $\times$ 80~mm$^2$ and thickness of 
20~mm.

\section{Fresnel Lenses}    
\label{sect:fresnel}
 {\index{Diffraction limited}}
Fresnel lenses have been extensively discussed and studied for use in astronomy at X-ray energies  ({\emph{e.g.} } 
\cite{1996SPIE.2805..224D,2004ApOpt..43.4845S,2004SPIE.5168..411G,
2006ExA....21..101B,2008SPIE.7011E..0UG, 
2012ApOpt..51.4638B,2018ApOpt..57.1857B} )
and missions exploiting them in that band  have been proposed
\cite{2008SPIE.7011E..0TS,2012SoPh..279..573D,2020SPIE11444E..7VK}.  
For a review see \cite{2010XROI.2010E..16S}. 
The circumstances in which such lenses offer diffraction limited resolution are discussed in  \ref{willingale}. 

Although the idea of \textcolor{green}{Fresnel  lenses} \index{Fresnel lenses}  for gamma-rays  was introduced at least 
as far back as 2001
\cite{2001A&A...375..691S,2002A&A...383..352S}.     
 when their potential for micro-arc-second imaging was pointed out, the idea has rested largely dormant. 
 It will be seen below that the main reason for this is the extremely long focal lengths of such lenses. A secondary reason is that although they offer effective areas far greater than any other technique, with a simple lens  the bandwidth over which this is achieved is narrow because of chromatic aberration. However, if those difficulties can be overcome, gamma-ray Fresnel lenses offer some unique possibilities. 
Like Laue lenses, Fresnel lenses provide a way of concentrating incoming gamma-rays onto a small, and hence low background, detector. A gamma-ray Fresnel lens could focus the flux incident on an aperture that could be many square metres  into a millimetre scale spot with close to 100\% efficiency. Moreover, such a lens would  also provide true imaging in the sense that there is a  one-to-one correspondence between incident direction and  positions in a focal plane. At photon energies above the limits of grazing energy optics no other technique can do this. Finally, the imaging can be diffraction-limited, which in the gamma-ray band with a metre scale aperture means sub-micro-arcsecond resolution.
What is more, even if missions employing them present challenges, gamma-ray Fresnel lenses are, \emph{per se}, low-technology items.

\begin{figure}[htbp]
\begin{center}
\includegraphics[trim = 20mm 35mm 10mm 35mm, clip, width=5cm]{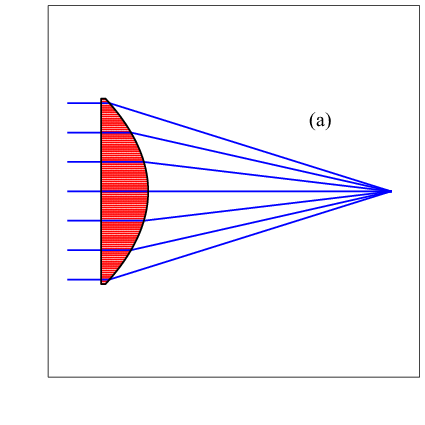}
\includegraphics[trim = 20mm 35mm 10mm 35mm, clip, width=5cm]{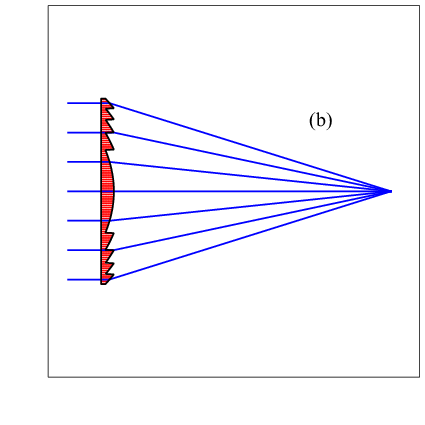}
\includegraphics[trim = 20mm 35mm 10mm 35mm, clip, width=5cm]{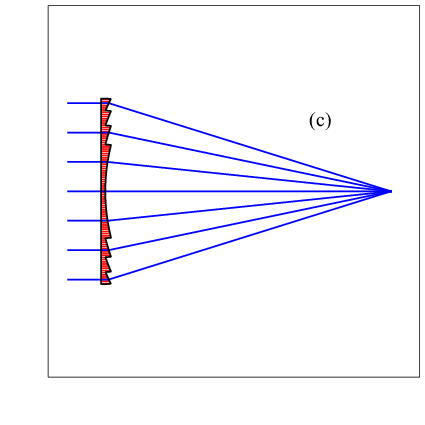}
\includegraphics[trim = 20mm 35mm 10mm 35mm, clip, width=5cm]{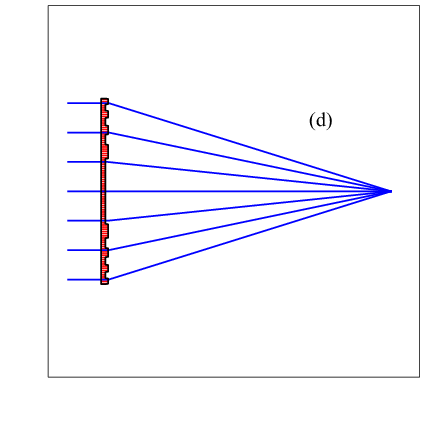}
\caption{(a) A conventional refractive lens for visible radiation (b) An equivalent Fresnel lens (c) A gamma-ray Fresnel lens (d) A gamma-ray Phase Zone Plate.}
\label{fig:1}
\end{center}
\end{figure}

A conventional refractive lens, operating for example in the visible band, focuses radiation  by introducing a radius-dependent delay such that radiation from different parts of the lens arrives at the focal spot with the same phase (Figure  \ref{fig:1}a).  A Fresnel lens \footnote{Strictly the term used should be ``Phase Fresnel Lens'' as the shorter form can also be used for stepped lenses in which coherence is not maintained between steps}   (Figure \ref{fig:1}b) achieves the same phase-matching by taking advantage of the fact that the phase of the incoming radiation never needs to be changed by more than $2\pi$. Consequently the maximum thickness of the lens can be reduced to  that necessary to produce a phase change of $2\pi$, a thickness termed here  $t_{2\pi}$.  It is usual to write the complex \textcolor{green}{refractive index} \index{Refractive index} as $ n = 1 - \delta - i\beta $ 
 where for gamma-rays  both $\delta$ and $\beta$ are small and positive. The imaginary component describes absorption and does not affect the phase so $t_{2\pi} = \lambda/\delta$ where $\lambda$ is the wavelength.   The fact that  $\delta$ is positive for gamma-rays means that  a converging lens has a concave profile and a Fresnel lens has the form illustrated in Figure  \ref{fig:1}c. It is a more efficient form of a  \textcolor{green}{`Phase Zone Plate`} \index{Phase Zone Plate} (Figure \ref{fig:1}d) in which the thickness profile has just two levels differing in phase shift by $\pi$. 
 
The parameter $\delta$ is given in terms 
of the atomic scattering factor $f_1(x,Z)$ discussed in Section \nameref{CrystalDiff}  by:

 \begin{equation}
\delta  = \frac{r_e\lambda^2}{2\pi}n_a f_1(x,Z)
\label{eq:delta}
\end{equation}
where $r_e$ is the classical electron radius and $n_a$ is the atomic density. For lenses of the type considered here $x$ is essentially zero.
Well above all absorption edges $f_1$ approaches the atomic number $Z$ and so is constant. Thus $\delta$ is proportional to $\lambda^2$ or inversely proportional to the square of the photon energy $E$. 

In principle in the region of the 1.022 MeV  threshold for pair production in the nuclear electric field and above, Delbr\"uck scattering should  be taken into consideration. Early reports of an unexpectedly large contribution to $\delta$ from this effect
 \cite{2012PhRvL.108r4802H} 
turned out to be mistaken \cite{2017PhRvL.118p9904H} 
but did lead to experimental confirmations of predicted gamma-ray refractive indices at energies up to 2 MeV \cite{2017PhRvA..95e3864G, 2017PhLA..381.3129K}. 
Like those from Delbr\"uck scattering, contributions from nuclear resonant scattering will also be negligible in most circumstances.

Although  $\delta$ is extremely small at gamma-ray energies, the wavelength is also extremely short, for example 1.24 pico-metres at 1 MeV.
Using $\delta$ from Eq. \ref{eq:delta} results in the following expression for $t_{2\pi}$ in terms of some example parameters
\begin{equation}
t_{2\pi} = \frac{\lambda}{\delta} = 2.98 \biggl(\frac{Z}{A}\biggr)^{-1}  \biggl(\frac{E}{1 \textrm{ MeV}} \biggr) \biggl(\frac{\rho}{1 \textrm{ g cm}^{-3}}\biggr)^{-1} \textrm{ mm}.
\end{equation}
Noting that for materials of interest $Z/A$ is in the range 0.4 to 0.5, this means that a gamma-ray Fresnel lens need have a thickness only of the order of millimetres.  
\begin{figure}[htbp]

\begin{center}
\includegraphics[trim = 0mm 0mm 0mm 0mm, clip, width=8cm]{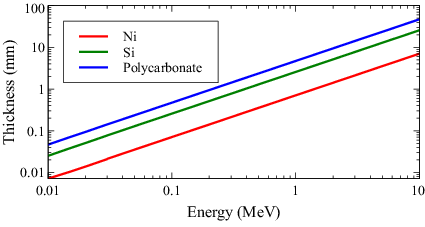}
\includegraphics[trim = 0mm 0mm 0mm 0mm, clip, width=8cm]{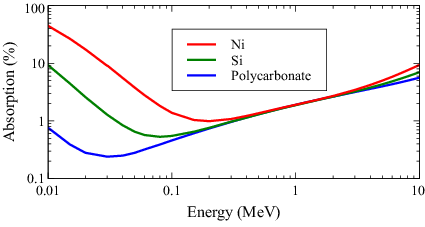}
\caption{(a) The thickness of some example materials needed to produce a $2\pi$ phase shift, plotted as a function of energy. (b) The mean absorption of a lens having as its maximum thickness that shown in (a). Absorption in any supporting substrate is not taken into account.}
\label{fig:2}
\end{center}
\end{figure}

The period of the sawtooth profile where it is lowest at the edge of the lens is a crucial parameter. Again in terms of example parameters, for a lens of diameter $d$ and focal length $F$ the minimum period is given by
\begin{equation}
p_{min} = 2\lambda \frac {F}{d} = 2.48 \biggl( \frac{E}{1\; \textrm{ MeV}}\biggr)^{-1} \biggl( \frac{F}{10^9 \textrm{ m}}\biggr) \biggl( \frac{d}{1\; \textrm{ m}}\biggr)^{-1} \textrm{ mm}.
\label{eqn:pmin} 
\end{equation}
The difficulties lie in the $F$ term. For reasonable values of $p_{min}$ extremely long focal lengths are required. 

$p_{min}$ is an important parameter in another respect. The 
PSF will be an Airy function with a FWHM of $1.03\lambda/d$, so the focal spot size using this measure will be given by
\begin{equation}
w = 1.03 \frac{F \lambda}{d} =  0.501 p_{min} 
\label{eqn:focalspot}
\end{equation}
That is to say the size of the focal spot is about half the period at the periphery of the lens. The corresponding angular resolution is $w/F$  and is given in {\bf micro-}arcseconds  ($\mu"$) by
\begin{eqnarray}
  \label{eqn:diff_lim1}
  \Delta \theta_d&=&   0.263  \left(    \frac {E}{1\: \textrm{MeV}} \right)^{-1}
  \left(    \frac {d}{1\: \textrm{m}} \right)^{-1}. 
    \label{eqn:diff_lim}
\end{eqnarray}
Thus gamma-ray Fresnel lenses have the potential to form images with an angular resolution better than available in any other waveband and would be capable of resolving, for example, structures on the scale of the event horizons of extra-galactic massive black holes. 

\subsection{Construction}
\label{sect:construct}

If the full potential of a Fresnel lens is to be realised, then Eq. \ref{eqn:focalspot} implies that $p_{min}$ should be  larger than the detector spatial resolution and so should be of the order of millimetres or more. Except at energies above 1 MeV the required thickness is also of the order of millimetres (Fig. \ref{fig:2}a) so high aspect ratios are not then needed. Almost any convenient material can be used -- the nuclei only serve to hold in place the cloud of electrons that produces the phase shifts!  Dense materials have the advantage of minimising thickness but also tend to be of higher $Z$ and hence more lossy at low and high energies. Over much of the gamma-ray band all reasonably low $Z$ materials have similarly low losses (Figure  \ref{fig:2}b).  Because the refractive index is so close to unity, dimensional tolerances are relatively slack. Using the Mar\'echal formula \cite{1982JOSA...72.1258M}, if the  \textit{rms} errors in the profile are kept within 3.5\% of the maximum lens thickness (assumed to be $t_{2\pi}$)  the loss in Strehl ratio (on-axis intensity) will be less than 5\%.  If the lens is assembled from segments then the  precision needed in their alignment is only at a similar level. Thus a range of  constructional technologies can be considered, including diamond point turning, vapour-phase deposition, photo-chemical etching and 3-d printing. 

\subsection{The focal length problem}
As argued above, if the  the best possible angular resolution is sought then detector consideration drive one to a $p_{min} \sim \textit{mm}$. When this is coupled with a requirement for a reasonable collecting area Eq. \ref{eqn:pmin} implies a focal length of the order of 
$10^5{\mbox{--}}10^6$ km. This sort of focal length  clearly demands \textcolor{green}{formation flying} \index{Formation flying} of two spacecraft, one carrying the lens and the other a focal plane detector. Minimising station-keeping propulsion requirement suggests locations near to a Sun-Earth Lagrangian point.
 
 The line of sight from detector to lens must be controlled with sufficient accuracy to keep the image on the detector, so the precision needed depends on the size of the detector. This could be from a few cm up to perhaps 1m. Changes in the direction of that line of sight need to be determined with an accuracy corresponding to the angular resolution aimed for, which could be sub-micro-arcsecond. 
 
 Detailed studies have been performed of missions with requirements that meet all of these requirements separately. Not all are met together in any single study,  but on the other hand the complexity of the instrumentation and spacecraft  was in each case much greater than that for a gamma-ray Fresnel telescope. 
 New Worlds Observer requires a 50~m star shade and a 4~m visible-light diffraction-limited telescope to be separated by $8\times 10^4$~km \cite{2009SPIE.7436E..06C}.  The transverse position of the telescope must be maintained to within a few cm. The  \textcolor{green}{MAXIM} {\index{MAXIM}} X-ray 
 interferometer mission calls for a fleet of 26 spacecraft distributed over an  area 1 km in diameter to be separated from a detector craft $2\times 10^4$ km away \cite{2003ExA....16...91C, 2005AdSpR..35..122C}. 
 The requirement for a diffraction-limited gamma-ray Fresnel lens mission to track 
 changes in the orientation of the configuration in inertial space at the sub-micro-arcsecond level are similar to the corresponding requirement for MAXIM, though only two spacecraft are needed. MAXIM studies envisaged a `super star tracker'. Such a device could locate a beacon on the lens spacecraft relative to a stellar reference frame. 
   In terms of separation distance, the requirement for $\sim 10^6$ km separation can be compared with the needs  of the LISA gravitational wave mission \cite{2021AdSpR..67.3868J} for which 3 spacecraft must each be separated from the others by $2.5\times 10^6$ km,  though in this case it is the rate of change of inter-spacecraft distance that requires strict monitoring rather than the orientation. 

\subsection{Effective area}
\label{sect:aeff} 

The focusing efficiency of a gamma-ray Fresnel lens will depend on the fraction of incoming radiation that passes unabsorbed through the lens and on the efficiency with which that radiation is concentrated into a focal spot. 

In Figure \ref{fig:2}b  the mean transmission of a lens with a maximum thickness of $t_{2\pi}$  and a typical profile is shown  for some example materials. 
Transmissions in excess of 95\% should generally be possible. For an Airy disc 84\% of the radiation falls within the central peak, that is to say within a radius equal to the resolution according to the Rayleigh criterion. As noted in Section \nameref{sect:construct}, 
allowing for profile errors with an rms level of 3.5\% \textit{rms} of the maximum height could reduce the Strehl ratio by 5\%. If the errors are random, the form of the PSF will be little changed and so the effective area will be reduced in proportion. Combining all these factors together indicates a focussing efficiency of about 75\%.

We take as a reference design a 2~m diameter lens made from polycarbonate with a nominal working energy of 500 keV and focal length $10^6$~km. 
The baseline active thickness is taken to be $t_{2\pi} = 1.15$~mm but absorption in a 2 mm substrate has been allowed for. Additional support against diaphragm mode vibrations would obviously be needed during launch but out of plane distortions during operation have little effect.
  
Allowing for the above factors,  an effective area over 20000~cm$^2$  should be attainable with a simple Fresnel lens having these parameters. The diffraction-limited angular resolution of such a lens would be 0.31~micro-arc-seconds. 
 
\subsection{Chromatic aberration}

Unfortunately the above indication of a possible effective area applies only at the particular energy and focal distance for which the lens has been designed. An important limit to the performance of a Fresnel lens is the chromatic nature of the imaging. The bandwidth over which the good focussing is achieved is very narrow. For even quite small deviations from the wavelength for which the profile has been designed the focal length changes and for a fixed detector position the focal spot rapidly becomes blurred.  The FWHM of the on-axis intensity as a function of energy is approximately
\begin{equation}
\frac{\delta \lambda}{\lambda} = \frac{\delta E}{E} = \frac{ 1.79}{ N_F}
\label{eqn:deltaE}
\end{equation}
where $N_F$ is the \textcolor{green}{Fresnel number}, \index{Fresnel number} $r^2/(f\lambda)$. When the lens thickness is $t_{2\pi}$  then $N_F$ is equal to twice the number of rings.

\begin{figure}[htbp]
\begin{center}
\hfill
\includegraphics[trim = 5mm 0mm 0mm 0mm, clip, width=58mm]{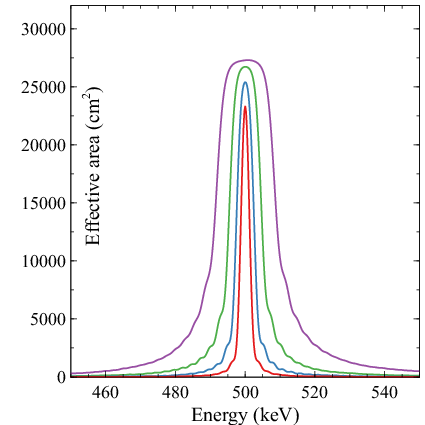}
\hfill
\includegraphics[trim = 5mm 0mm 0mm 0mm, clip, width=58mm]{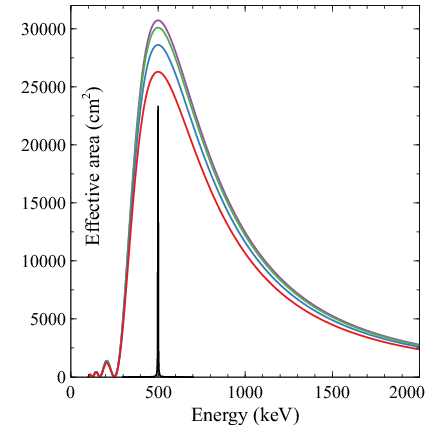}
\hspace{5mm}
\caption{(a) The effective area of a Fresnel lens with the parameters described in Table~\ref{table:1}. The curves assume that the flux is collected from a circular region with a radius 1, 2, 4 and 8 times the 1.5 mm Rayleigh width of a diffraction limited response (narrowest peak to widest).    The FWHM increases from 2.8 to 17.7 keV.  (b) As (a) but assuming that at each energy the configuration is refocussed by moving  the detector to the focal plane for that energy. The narrowest peak from (a) is reproduced for comparison.}
 \label{fig:3}
\end{center}
\end{figure}

If the flux within a focal spot the size of the ideal Airy peak is used as a measure of the response, then the bandwidth is somewhat larger, with the numerical factor in Eq. \ref{eqn:deltaE} increased to about 2.95, and if
the flux from a larger detector region is  accepted, then the bandwidth  increases further as seen in Figure \ref{fig:3}. The improvement will be at the expense of poorer angular resolution and increased detector background. Note that the diffraction-limited angular resolution will anyway only be available if  the detector energy resolution is good enough to select only those photons within $\Delta E$. For a High Purity Germanium detector (see Volume 2 of this work,  \ref{germanium} ) then
 $\frac{\delta E}{E} \sim 4\times 10^{-3}$ at 500 keV, effectively setting a limit of about 700 on the useful Fresnel Number of a  diffraction limited telescope using a simple gamma-ray Fresnel lens.
 
Over a much wider band, good focussing can be recovered by adjusting the 
detector position (Figure \ref{fig:3})  but only radiation within the 
narrow bands can be focussed efficiently for any given position. 
There is a direct analogy with the tunable Laue lenses described 
in \nameref{sect:tunable}, but for a Fresnel lens no adjustment is required 
to the lens, only a change in focal distance.

For visible and IR radiation it has been shown possible to increase the bandpass of Fresnel lenses by correcting the chromaticity of one diffractive lens with that of another one of opposite power, but such systems cannot produce a real image as is required for a gamma-ray telescope \cite{1976ApOpt..15..542B, 2010XROI.2010E..16S}.
Similarly schemes used in those wavebands to make "Multiorder" or "Harmonic" Fresnel lenses by increasing the maximum thickness of the lens so that it  becomes a (different) integer multiple of $t_{2\pi}$ at each wavelength \cite{1995ApOpt..34.2462F, 1995ApOpt..34.2469S} do not work in the gamma-ray band. A small part of the surface of such a lens can be regarded as a diffraction grating blazed to direct radiation into the appropriate order, but as  $t_{2\pi}$  for gamma-rays is almost universally proportional to $\lambda^{-1}$, the blaze angle cannot be correct at different energies.

However various possibilities do exist for alleviating the chromaticity problem:

\begin{enumerate}

\item{ \textcolor{green}{`Axilenses'} \index{Axilenses}  in which the pitch is varied in ways other than the $r^{-1}$ variation of a regular Fresnel lens. This can produce an enlarged bandwidth at the expense of efficiency \cite{2009SPIE.7437E..0JS}. The integral over energy of the effective area is close to that of a corresponding classical Fresnel lens.
}
\item{\textcolor{green}{Achromatic doublets.} \index{Achromatic doublets}
While  the focal length of a  diffractive lens for X-rays or gamma-rays is proportional to $E$,  that of a refractive lens is even more strongly dependent on energy, being proportional to $E^2$.  In various contexts 
\cite{2002A&A...383..352S, 2003SPIE.4851..599G, 2003Natur.424...50W}  
it has been pointed out that  consequently  diffractive and refractive lenses for which the focal lengths are
\begin{equation}
 f_d = \biggl(\frac{E}{E_0}\biggr) \frac{f_0}{2} \hspace{10mm}  f_r = -\biggl(\frac{E}{E_0}\biggr)^2 {f_0}
\end{equation}
 can be combined to form an achromatic doublet for which the combined focal length is to first order independent of wavelength.  In the gamma-ray band the absorption losses for a full refractive lens with useful parameters are likely to be prohibitive but steps that are many times $t_{2\pi}$ can be introduced. This leads to a lens having a profile corresponding to the combination of those shown in Figure \ref{fig:4}(a) and (b). Fig \ref{fig:4}(c) is a zoom on parts of the fine structure in the diffractive component, (b). At a given radius only the total thickness is important, so the two components can be separate as shown, or back to back in s single component, or the diffractive profile can be superimposed on that of the refractive one. With such  lens focusing then be achieved at a number of wavelengths   over an extended band (Figure \ref{fig:5}(a)). The peak effective area is less than that for a single lens but the integral can be several times higher.  The mass of the lens will be many times that of a simple lens and the finest scale of the structure must be a factor of two smaller. 
}
\begin{figure}[htbp]
\begin{center}
\includegraphics[trim = 5mm 40mm 0mm 0mm, clip, width=80mm]{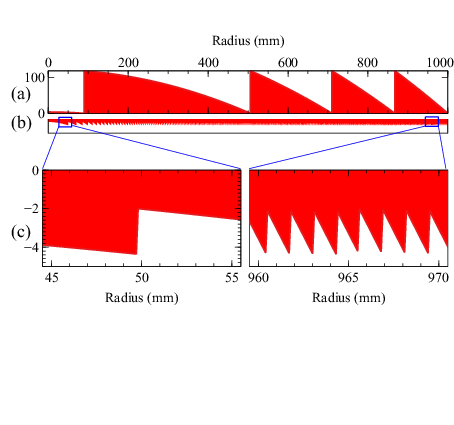} 
\hspace{5mm}
\caption{Profile of an achromatic doublet. (a) The refractive component (b) The diffractive component (Fresnel lens) component (c) Zooms on sample regions of (b).  }
\label{fig:4}
\end{center}
\end{figure}

\begin{figure}[htbp]
\begin{center}
\hfill
\includegraphics[trim = 5mm 0mm 0mm 0mm, clip, width=58mm]{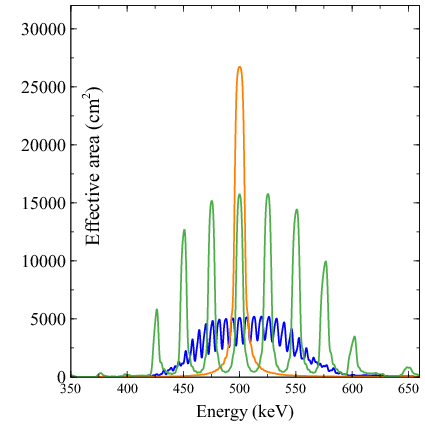}
\hfill
\includegraphics[trim = 5mm 0mm 0mm 0mm, clip, width=58mm]{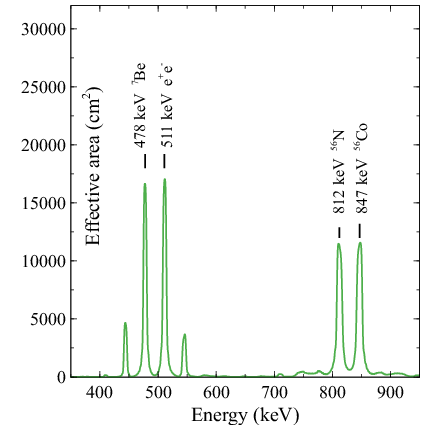}
\caption{(a)  The effective area of `achromatic'  doublets in which a simple  converging Fresnel diffractive lens is combined with a diverging refractive one. The refractive component is assumed to be stepped at points where the thickness would exceed either 20 (green) or 80 (blue) times $t_{2\pi}$ at the nominal energy of 500 keV. The flux is assumed to be collected from a spot  4 times the diameter of the central peak of a diffraction limited response. The corresponding curve from Figure \ref{fig:3} is reproduced (red line) for comparison.  (b) The corresponding  effective area for a multiwavelength lens  optimised for response at 4 energies of astrophysical interest.  }
\label{fig:5}
\end{center}
\end{figure}

\item{\textcolor{green}{Multi-wavelength  lenses}. \index{Multi-wavelength lenses} 
Recently, a number of papers have been published describing the design of `achromatic diffractive lenses' for UV/visible/IR radiation (e.g. \cite{BanerjiSWIR}). The approach used is to optimise the thickness of a number of concentric rings so that the imaging performance over a number of wavelengths is as good as possible based of some figure of merit. The same technique can be used to design gamma-ray lenses.  At first sight the reported performance of the IR lens designs is remarkably good with, for example, 91\% `average efficiency' reported over wavelengths spanning a factor of 2.  On detailed examination, though, the efficiency is low except  at a few specific wavelengths for which the design was optimised and analysis suggests that even at these wavelengths the efficiency may have been overestimated \cite{2021Optic...8..834E}.  

If the maximum thickness is restricted to be $\sim t_{2\pi}$ then the integrated effective area of a lens designed in this way is found to be similar to that of a simple Fresnel lens. If larger thicknesses are allowed then the performance and limitations are similar to those for an achromat of the same total thickness but there is the possibility of choosing the energies at which the performance is optimised.
An example is illustrated in Figure~\ref{fig:5}~(b) which shows the effective area of a lens designed to work simultaneously at the energies of 4 astrophysical gamma-ray lines. The  approach used is basically that described by Doskolovich  \emph{et al.} \cite{2020OExpr..2811705D} which maximises the Strehl ratio averaged over the design energies.  However an additional step was performed in which different  target phases were tried and that selected which  resulted in the highest value for the minimum over the four energies of the effective area, thus avoiding solutions in which one dominated.
}

\begin{table}[ht]
\caption{Parameters of example lens designs. All the designs have the following in common - Nominal energy: 500 keV, Diameter: 2 m, Focal distance: $10^6$ km, Fresnel Number: 400,  Diffraction limited angular resolution 0.3 micro-arc-seconds, Material: Polycarbonate.  A substrate of 2 mm is allowed for in the mass and in calculating the effective area but not included in the 'active thickness'.  The effective areas quoted are those for the lens and do not include detector efficiency. Those quoted here assume that the flux is collected from a 12 mm diameter region in the image plane.  \emph{rms} figure errors of 3.5\% of $t_{2\pi}$ have been allowed for. }
\label{table:1}
\begin{center}
\begin{tabular}{|c|c|c|c|c|c|c|}
\hline
\multirow{2}{*}{Type}    & \multirow{2}{*}{Figure}  &  \multicolumn{2}{c|}{Active thickness $t$}    & Mass    &   \multicolumn{2}{c|}{Effective Area} \\ \cline{3-4}\cline{6-7}
                                    &                    &  mm                  &    $t/t_{2\pi}$   &    kg      &   Peak (cm$^2$)  & Integral ( cm$^2$ MeV ) \\
 \hline                         
Simple Fresnel            & \ref{fig:3}a &      2.4                 &           1             &     12         &     26735              &        331                      \\                         
Achromatic Doublet     & \ref{fig:5}a(i)  &    50                &          21           &    100        &   15768              &         720                     \\   
       "                             & \ref{fig:5}a(ii)  &    191             &          81           &    332       &      5224               &         440                    \\   
Multiwavelength           &  \ref{fig:5}b     &      24              &          10            &        52      &     17064               &         642                     \\   
\hline
\end{tabular}
\end{center}
\label{default}
\end{table}

\end{enumerate}
\newpage

\section{Detector issues for focused gamma-rays}

A discussion of  detectors for gamma-ray astronomy can be found in Volume \ref{germanium} of this work. 
Here we present some specific considerations for detectors for focusing gamma-ray telescopes.

\begin{itemize}
\item  {\bf Low Background.} 
Gamma-ray observations are almost  always limited by background from diffuse sky emission 
and events   
arising directly or indirectly from cosmic-rays. 
It is imperative to suppress these as far as possible.
Shielding is heavy and even active shielding is imperfect, some background actually being created by the shielding itself.  
\item {\bf Sensitivity to Photon Arrival Direction.}
A key method of background reduction in the low energy gamma-ray band involves extracting information about the direction from which a photon arrived. Where the first interaction of a photon in the detector is a Compton scattering and the 3-d locations of that and subsequent interactions can be recorded a Compton kinematics analysis allows constraints to be placed on the arrival direction. Although there are ambiguities and in the MeV range the precision is limited to a few degrees, background can be drastically reduced by selecting only events whose direction is consistent with having passed through the lens. Obviously such analysis is not possible for all events, but it is possible to identify a subset of low detection efficiency, very low background, events and give these a high weight during analysis. 
\item{\bf Position Resolution.}  
 In an imaging system it is strictly the position where a
photon first interacts in the detector that is important, though the location of the centroid of
all energy deposits may provide an adequate approximation.
Even when the instrument
is used as a flux-collector the position information can be used to define a region of the detector
within which events are counted or, better, to attribute  to each event a weight  depending 
 on the noise level and the expected response at its location. The latter amounts to 'PSF fitting'.
Of course Compton kinematics analysis requires
recording the positions in 3-d of subsequent interactions as well as the first.

\item {\bf Spatial extent}
For a Laue lens a large detector will  allow all of the events in the broad wings of the PSF to be captured. With the long focal lengths associated with Fresnel lenses, fields of view tend to be very small and are limited by how large a detector is feasible. In either case a large detector can provide regions that can be used for contemporaneous  background estimation, obviating or reducing the need for 'off-source' observations
\item {\bf Energy Resolution.}
As gamma-ray lenses operate over a limited band, good energy resolution reduces the amount 
of background accompanying the signal. 
Fresnel lenses, particularly those that are large or have less extreme focal lengths, suffer from severe chromatic aberration so this is crucial where they are used.
More generally, insufficient 
energy resolution will degrade the angular resolution of the Compton kinematic 
reconstruction and so increase the background.

\item{\bf Detection efficiency}
Detection efficiency is a major problem in the MeV region. The photons are very 
penetrating and a non-negligible fraction of photons traverse the detector without interacting at all.
\end{itemize}

The constraints resulting from these objectives are often conflicting. A dense detector is 
desirable to maximise efficiency, but  for Compton reconstruction low density materials 
are favoured. Compton reconstruction requires measuring position and  the energy deposits at multiple sites but on the other hand electronic read-out noise has least impact on energy resolution if all of the charge is fed to a single preamplifier.

High purity Germanium detectors provide good efficiency and currently the best energy resolution. Germanium based Compton cameras are well advanced and demonstrated 
\cite{Tomsick2019}. Cadmium-Zinc-Telluride (CZT) elements have an energy resolution that is only slightly inferior, do not need cooling and can be assembled in large arrays.
More complex detectors using a 3d-imaging Silicon scattering column surrounded by a
3d-imaging CZT-shield absorbing the scattered photons  may reach a higher efficiency, but are currently at a development stage \cite{kuvvetli05}.  

\section{Conclusions}
\label{conclusions}

Diffractive gamma-ray optics is a field yet to be 
explored and exploited. Nevertheless, Laue and Fresnel 
lenses offer unique possibilities. The examples shown 
in this review illustrate both their capabilities and 
their limitations. An important consideration in choosing 
a lens design is the expected signal and how it compares 
with the likely event rate due to background in the detector. 
Increasing the effective area is not necessarily an advantage 
if the background noise increases. For photon-starved 
observations of continuum sources, though, large 
integrated effective areas are needed. 
For applications in which lower effective area or 
poorer resolution are adequate, reduced diameter or 
shorter focal lengths can be considered. In some cases an 
array of smaller lenses can provide the best solution.

Observational MeV astrophysics is still in its infancy. 
We only know of a handful of low energy gamma-ray point sources. 
This is even less than the number known at 
100~MeV and above from the SAS-2 and COS-B missions in the 1970s and 1980s. 
Today high energy gamma-ray sources are  counted in 
the thousands. For MeV astrophysics we now await 
results 
from the COSI mission \cite{tomsick19}, which 
in a few years will give us a much clearer 
picture of the sky at a few MeV,  that 
will be very important in designing future gamma-ray telescopes.

\section{Cross References}

\begin{enumerate}
\item {Willingale, Richard. Handbook of X-ray and Gamma-ray Astrophysics,
 Volume 1: X-ray Experimental Techniques and Missions, ``Diffraction-limited optics and techniques" \label{willingale}}

\item{ TBD, Handbook of X-ray and Gamma-ray Astrophysics, 
Volume 2: Gamma-ray Experimental Techniques, Observatories, and Missions, ``Germanium detectors for gamma-ray astronomy" \label{germanium}}

\end{enumerate}

\bibliographystyle{ws-book-van}
\bibliography{Laue_Fresnel_arxiv}

\printindex

\end{document}